\documentclass[12pt]{article}

\usepackage{amsmath,amsthm,amssymb,euscript,bbold,epsf,epsfig}
\usepackage{array}

\def\DottedCircle{
\qbezier[4](0.966,-0.259)(1.04,0)(0.966,0.259)
\qbezier[4](0.966,0.259)(0.897,0.518)(0.707,0.707)
\qbezier[4](0.707,0.707)(0.518,0.897)(0.259,0.966)
\qbezier[4](0.259,0.966)(0,1.04)(-0.259,0.966)
\qbezier[4](-0.259,0.966)(-0.518,0.897)(-0.707,0.707)
\qbezier[4](-0.707,0.707)(-0.897,0.518)(-0.966,0.259)
\qbezier[4](-0.966,0.259)(-1.04,0)(-0.966,-0.259)
\qbezier[4](-0.966,-0.259)(-0.897,-0.518)(-0.707,-0.707)
\qbezier[4](-0.707,-0.707)(-0.518,-0.897)(-0.259,-0.966)
\qbezier[4](-0.259,-0.966)(0,-1.04)(0.259,-0.966)
\qbezier[4](0.259,-0.966)(0.518,-0.897)(0.707,-0.707)
\qbezier[4](0.707,-0.707)(0.897,-0.518)(0.966,-0.259)
}
\def\FullCircle{
\thicklines
\put(0,0){\circle{2}}
}
\def\Endpoint[#1]{
\ifcase#1
\put(1,0){\circle*{0.15}}
\or\put(0.866,0.5){\circle*{0.15}}
\or\put(0.5,0.866){\circle*{0.15}}
\or\put(0,1){\circle*{0.15}}
\or\put(-0.5,0.866){\circle*{0.15}}
\or\put(-0.866,0.5){\circle*{0.15}}
\or\put(-1,0){\circle*{0.15}}
\or\put(-0.866,-0.5){\circle*{0.15}}
\or\put(-0.5,-0.866){\circle*{0.15}}
\or\put(0,-1){\circle*{0.15}}
\or\put(0.5,-0.866){\circle*{0.15}}
\or\put(0.866,-0.5){\circle*{0.15}}
\fi}
\def\Arc[#1]{
\thicklines			
\ifcase#1
\qbezier[25](0.966,-0.259)(1.04,0)(0.966,0.259)
\or
\qbezier[25](0.966,0.259)(0.897,0.518)(0.707,0.707)
\or
\qbezier[25](0.707,0.707)(0.518,0.897)(0.259,0.966)
\or
\qbezier[25](0.259,0.966)(0,1.04)(-0.259,0.966)
\or
\qbezier[25](-0.259,0.966)(-0.518,0.897)(-0.707,0.707)
\or
\qbezier[25](-0.707,0.707)(-0.897,0.518)(-0.966,0.259)
\or
\qbezier[25](-0.966,0.259)(-1.04,0)(-0.966,-0.259)
\or
\qbezier[25](-0.966,-0.259)(-0.897,-0.518)(-0.707,-0.707)
\or
\qbezier[25](-0.707,-0.707)(-0.518,-0.897)(-0.259,-0.966)
\or
\qbezier[25](-0.259,-0.966)(0,-1.04)(0.259,-0.966)
\or
\qbezier[25](0.259,-0.966)(0.518,-0.897)(0.707,-0.707)
\or
\qbezier[25](0.707,-0.707)(0.897,-0.518)(0.966,-0.259)
\fi}
\def\DottedArc[#1]{
\ifcase#1
\qbezier[4](0.966,-0.259)(1.04,0)(0.966,0.259)
\or
\qbezier[4](0.966,0.259)(0.897,0.518)(0.707,0.707)
\or
\qbezier[4](0.707,0.707)(0.518,0.897)(0.259,0.966)
\or
\qbezier[4](0.259,0.966)(0,1.04)(-0.259,0.966)
\or
\qbezier[4](-0.259,0.966)(-0.518,0.897)(-0.707,0.707)
\or
\qbezier[4](-0.707,0.707)(-0.897,0.518)(-0.966,0.259)
\or
\qbezier[4](-0.966,0.259)(-1.04,0)(-0.966,-0.259)
\or
\qbezier[4](-0.966,-0.259)(-0.897,-0.518)(-0.707,-0.707)
\or
\qbezier[4](-0.707,-0.707)(-0.518,-0.897)(-0.259,-0.966)
\or
\qbezier[4](-0.259,-0.966)(0,-1.04)(0.259,-0.966)
\or
\qbezier[4](0.259,-0.966)(0.518,-0.897)(0.707,-0.707)
\or
\qbezier[4](0.707,-0.707)(0.897,-0.518)(0.966,-0.259)
\fi}
\def\Chord[#1,#2]{
\thinlines
\ifnum#1>#2\Chord[#2,#1]
\else\ifnum#1<#2
\ifcase#1
\ifcase#2
\or\qbezier(1,0)(0.516,0.138)(0.866,0.5)
\or\qbezier(1,0)(0.45,0.26)(0.5,0.866)
\or\qbezier(1,0)(0.327,0.327)(0,1)
\or\qbezier(1,0)(0.179,0.311)(-0.5,0.866)
\or\qbezier(1,0)(0.0536,0.2)(-0.866,0.5)
\or\put(1, 0){\line(-2, 0){2}}
\or\qbezier(1,0)(0.0536,-0.2)(-0.866,-0.5)
\or\qbezier(1,0)(0.179,-0.311)(-0.5,-0.866)
\or\qbezier(1,0)(0.327,-0.327)(0,-1)
\or\qbezier(1,0)(0.45,-0.26)(0.5,-0.866)
\or\qbezier(1,0)(0.516,-0.138)(0.866,-0.5)
\fi
\or\ifcase#2\or
\or\qbezier(0.866,0.5)(0.378,0.378)(0.5,0.866)
\or\qbezier(0.866,0.5)(0.26,0.45)(0,1)
\or\qbezier(0.866,0.5)(0.12,0.446)(-0.5,0.866)
\or\qbezier(0.866,0.5)(0,0.359)(-0.866,0.5)
\or\qbezier(0.866,0.5)(-0.0536,0.2)(-1,0)
\or\put(0.866, 0.5){\line(-5, -3){1.73}}
\or\qbezier(0.866,0.5)(0.146,-0.146)(-0.5,-0.866)
\or\qbezier(0.866,0.5)(0.311,-0.179)(0,-1)
\or\qbezier(0.866,0.5)(0.446,-0.12)(0.5,-0.866)
\or\qbezier(0.866,0.5)(0.52,0)(0.866,-0.5)
\fi
\or\ifcase#2\or\or
\or\qbezier(0.5,0.866)(0.138,0.516)(0,1)
\or\qbezier(0.5,0.866)(0,0.52)(-0.5,0.866)
\or\qbezier(0.5,0.866)(-0.12,0.446)(-0.866,0.5)
\or\qbezier(0.5,0.866)(-0.179,0.311)(-1,0)
\or\qbezier(0.5,0.866)(-0.146,0.146)(-0.866,-0.5)
\or\put(0.5, 0.866){\line(-3, -5){1}}
\or\qbezier(0.5,0.866)(0.2,-0.0536)(0,-1)
\or\qbezier(0.5,0.866)(0.359,0)(0.5,-0.866)
\or\qbezier(0.5,0.866)(0.446,0.12)(0.866,-0.5)
\fi
\or\ifcase#2\or\or\or
\or\qbezier(0,1.)(-0.138,0.516)(-0.5,0.866)
\or\qbezier(0,1.)(-0.26,0.45)(-0.866,0.5)
\or\qbezier(0,1.)(-0.327,0.327)(-1,0)
\or\qbezier(0,1.)(-0.311,0.179)(-0.866,-0.5)
\or\qbezier(0,1.)(-0.2,0.0536)(-0.5,-0.866)
\or\put(0, 1){\line(0, -2){2}}
\or\qbezier(0,1.)(0.2,0.0536)(0.5,-0.866)
\or\qbezier(0,1.)(0.311,0.179)(0.866,-0.5)
\fi
\or\ifcase#2\or\or\or\or
\or\qbezier(-0.5,0.866)(-0.378,0.378)(-0.866,0.5)
\or\qbezier(-0.5,0.866)(-0.45,0.26)(-1,0)
\or\qbezier(-0.5,0.866)(-0.446,0.12)(-0.866,-0.5)
\or\qbezier(-0.5,0.866)(-0.359,0)(-0.5,-0.866)
\or\qbezier(-0.5,0.866)(-0.2,-0.0536)(0,-1)
\or\put(-0.5, 0.866){\line(3, -5){1}}
\or\qbezier(-0.5,0.866)(0.146,0.146)(0.866,-0.5)
\fi
\or\ifcase#2\or\or\or\or\or
\or\qbezier(-0.866,0.5)(-0.516,0.138)(-1,0)
\or\qbezier(-0.866,0.5)(-0.52,0)(-0.866,-0.5)
\or\qbezier(-0.866,0.5)(-0.446,-0.12)(-0.5,-0.866)
\or\qbezier(-0.866,0.5)(-0.311,-0.179)(0,-1)
\or\qbezier(-0.866,0.5)(-0.146,-0.146)(0.5,-0.866)
\or\put(-0.866, 0.5){\line(5, -3){1.73}}
\fi
\or\ifcase#2\or\or\or\or\or\or
\or\qbezier(-1,0)(-0.516,-0.138)(-0.866,-0.5)
\or\qbezier(-1,0)(-0.45,-0.26)(-0.5,-0.866)
\or\qbezier(-1,0)(-0.327,-0.327)(0,-1)
\or\qbezier(-1,0)(-0.179,-0.311)(0.5,-0.866)
\or\qbezier(-1,0)(-0.0536,-0.2)(0.866,-0.5)
\fi
\or\ifcase#2\or\or\or\or\or\or\or
\or\qbezier(-0.866,-0.5)(-0.378,-0.378)(-0.5,-0.866)
\or\qbezier(-0.866,-0.5)(-0.26,-0.45)(0,-1)
\or\qbezier(-0.866,-0.5)(-0.12,-0.446)(0.5,-0.866)
\or\qbezier(-0.866,-0.5)(0,-0.359)(0.866,-0.5)
\fi
\or\ifcase#2\or\or\or\or\or\or\or\or
\or\qbezier(-0.5,-0.866)(-0.138,-0.516)(0,-1)
\or\qbezier(-0.5,-0.866)(0,-0.52)(0.5,-0.866)
\or\qbezier(-0.5,-0.866)(0.12,-0.446)(0.866,-0.5)
\fi
\or\ifcase#2\or\or\or\or\or\or\or\or\or
\or\qbezier(0,-1.)(0.138,-0.516)(0.5,-0.866)
\or\qbezier(0,-1.)(0.26,-0.45)(0.866,-0.5)
\fi
\or\ifcase#2\or\or\or\or\or\or\or\or\or\or
\or\qbezier(0.5,-0.866)(0.378,-0.378)(0.866,-0.5)
\fi\fi\fi\fi}
\def\FullChord[#1,#2]{
\Endpoint[#1]
\Endpoint[#2]
\Arc[#1]
\Arc[#2]
\Chord[#1,#2]
}
\def\EndChord[#1,#2]{
\Endpoint[#1]
\Endpoint[#2]
\Chord[#1,#2]
}
\def\Picture#1{
\begin{picture}(2,1)(-1,-0.167)
#1
\end{picture}
}
\def\DottedChordDiagram[#1,#2]{
\Picture{\DottedCircle \FullChord[#1,#2]}
}
\newcommand{\abb}{\addtolength{\belowdisplayskip}{\belowdisplayskip}}
\newcommand{\Str}{{\rm Str}}
\newcommand{\Tr}{{\rm Tr}}
\newcommand{\ssk}{{\hspace{3pt}}}

\newcommand{\ccn}{ {{\cal{C}}_n}  }
\newcommand{\bp}{  \begin{pmatrix} }
\newcommand{\ep}{  \end{pmatrix} }
\newcommand{\pt}{ \partial }
\newcommand{\nnm}{ \nonumber }

\newcommand{\cLo}{ {\cal{L}}_{1} }
\newcommand{\cEo}{ {\cal{E}}_{1} }
\newcommand{\cLt}{ { \cal{L}}_{2} }
\newcommand{\cLz}{ { \cal{L}}_{0} }
\newcommand{\cEz}{ { \cal{E}}_{0} }
\newcommand{\cEt}{ { \cal{E}}_{2} }
\newcommand{\cO}{ {\cal{O}} }

\begin{document}

\begin{flushright}
QMUL-PH-04-04\\
{\tt hep-th/0405256}
\end{flushright}
\vskip0.1truecm
\begin{center}
\vskip 1truecm {\Large\bf
 Resolving  brane collapse with 1/N corrections
 in non-Abelian DBI
} \vskip 1truecm
{\large S Ramgoolam, B Spence and S Thomas${}^{\dagger}$\\
Department of Physics\\
Queen Mary, University of London\\
Mile End Road\\
London E1 4NS UK\\
}
\vskip 2truemm
\end{center}

\vskip 1truecm
\begin{center}
{\bf \large Abstract}
\end{center}
A collapsing spherical D2-brane carrying magnetic flux can be
described in the region of small radius in a dual zero-brane
picture using Tseytlin's proposal for a non-Abelian Dirac-Born-Infeld
action for $N$ D0-branes. A standard large $N$ approximation of
the D0-brane action, familiar from the Myers dielectric effect,
gives a time  evolution which agrees with the Abelian D2-brane
Born-Infeld equations which describe a D2-brane collapsing to zero
size. The first $1/N$ correction from the symmetrised trace
prescription in the zero-brane action leads to a class of
classical solutions where   the minimum radius of a collapsing
D2-brane is lifted away from zero. We discuss the  validity of
this approximation to the zero-brane action in the region of the
minimum, and explore higher order $1/N$ corrections as well as an
exact finite $N$ example. The $1/N$ corrected Lagrangians and the
finite $N$ example have an  effective mass squared which becomes
negative in some regions of phase space. We discuss the physics of
this tachyonic behaviour.

\vfill

\begin{flushright}
{\it ${}^{\dagger}${\{s.ramgoolam, w.j.spence, s.thomas\}@qmul.ac.uk}\\
}
\end{flushright}
\newpage


\section{Introduction}

There has been a lot of work recently on time dependence
in string theory \cite{seni,gutstro,senii,cc,lms,hopo,dupi}.
In this paper we will study a simple time dependent
system of a  spherical D2-brane  carrying magnetic flux.

D2-branes in type IIA string theory with magnetic flux
on the world-volume carry zero-brane charge
\cite{towns,dougbb}.
 We consider
such spherical D2-branes and their time evolution as described by
the 2+1 dimensional Born-Infeld action. A closely related problem
is  the time dependence of a spherical M2-brane of M-theory. When
considered in the context of Matrix Theory \cite{bfss,kata}, we
are led to add momentum along the eleventh direction, which
amounts in the IIA picture to having D2-branes with magnetic flux.
Other closely related systems include the microscopic description of 
giant gravitons \cite{janloz} 
and  the D3-brane bion where the
world-volume of the D3 stretches out into a D1-funnel described by
an excitation of the scalars in the D3-world-volume
\cite{calmal,gibbons}. Magnetic fluxes in the D3-worldvolume allow
the existence of a BPS solution describing such funnels. The dual
description in terms of the D1-worldvolume has been studied in
detail in \cite{cmtbcore}, one general lesson being that the D1
and D3 descriptions agree at large $N$.
 Whereas the D1-D3 funnel system involves
a spherically symmetric solution of the D1-world-volume
with the radius $ R( \sigma)$ of the spherical cross-section of the funnel
having a non-trivial dependence on the spatial coordinate
$\sigma$, the D0-D2 system has a time dependent radius $ R(t)$
of the spherical D2-brane.

In a similar spirit to \cite{cmtbcore} we may   look for the
collapsing D2-brane in terms of the D0-branes. The dualities
between descriptions of the same physics from two points of view,
of  a lower dimensional brane and a higher dimensional brane, can
thus be explored in a time-dependent context. The D0-brane
effective world-volume Yang Mills theory neglects stringy
excitations. This is a valid approximation when the separation of
the zero branes is less than the string length (branes as
short distance probes in string theory were studied in detail 
in \cite{dkps}).
 For a spherical
system of $N$ D0-branes this means that the radius $R$ of the
sphere obeys $ R < l_s \sqrt N $. The semiclassical world-volume
D2-brane action is expected to be valid for $ R > l_s $. There is
a large overlap of regimes of validity at large $N$. Indeed we
will find in Section 2 that the
non-Abelian Born-Infeld type action of the D0-branes
(written down in
 \cite{tseyt} and developed to describe
 the dielectric effect in  \cite{myers}) gives at large $N$
  the same equation for the time evolution of $R$ as
the one obtained from the D2-brane action.

In Section 3, we consider $ 1/N$ corrections to the equation for
$R$. This requires some combinatorics of symmetrised traces. We
give two methods for computing these corrections. The first is
based on evaluation of a symmetrised string of $su(2)$ generators
on the highest weight state of an irreducible representation.
 The second method
maps the problem of expressing the symmetrised generators in terms
of powers of the Casimir, into a combinatorics of chord diagrams
of the kind that appear in knot theory.

In Sections 4 and 5 we study the leading $1/N$ correction.
We consider the energy as a function of
$( r , s )$, the dimensionless position and velocity variables.
  For convenience
many formulae are expressed in terms of the variables $ ( \gamma ,
U ) $ where $ \gamma = ( 1  - s^2 )^{-{1 \over 2 }}$ is the
standard Lorentz factor of special relativity, and $ U = ( 1 + r^4
)^{1 \over 2 }$. At leading order all the formulae for Lagrangian,
energy, momentum, etc, are the standard ones of special
relativity, with $U$ playing the role of a position dependent
mass. The $1/N$ corrections give an interesting modification of
these standard formulae. The physical
consequence of the correction is that while a spherical brane
starting at rest at a large radius collapses to zero radius for a
range of radii, it cannot do so if the initial radius is too
large. These large collapsing membranes
 bounce from a minimal radius, which depends on the
initial radius. This is a somewhat exotic bounce where the
velocity $\dot {\bf r} $ is discontinuous, as in reflection at a
hard wall, and in addition the fate of the
membrane after the bounce is not uniquely determined, but rather
there are two possible outcomes. These conclusions follow from an
analysis of the energy function in $(r,s)$ space and the existence
of an extremum where $ { \partial E \over \partial s } = 0 $ and $
s \ne 0 $. We briefly discuss quantum mechanics near this
extremum.

In Section 6 we consider higher order corrections in the  $1/N$
expansion, explaining which properties of the solutions are
preserved as we include these. In Section 7 we consider the exact
finite $N$ evaluation of the symmetrised trace for the case of
spin $1/2$. We find explicit forms of the energy function in terms
of hypergeometric functions and discuss the dependence upon
position and velocity. In Section 8 we discuss regimes of validity
and the effects of higher order terms in general, paying attention
to the behaviour of the acceleration and effective mass, and the
appearance of tachyonic modes. Finally, in Section 9 we present a
summary and outlook.


\section{Time dependent solutions: Born-Infeld }

Consider the action describing $N$ D0-branes
\begin{equation}\label{dzero}
S_0 = -T_0 \int dt ~~ Str \sqrt{ (I_N - \lambda^2 \pt_t \Phi^i {Q}^{-1}_{ij}
           \pt _t \Phi^j )}\sqrt{{\rm det}\,Q^{ij} }
\end{equation}
where $\Phi^i, i = 1..9 $ are a set of $N\times N$ matrices and $Q^{ij} = I_N \delta^{ij} +
i \lambda [\Phi^i , \Phi^j] $. $T_0$ is the tension, $\lambda= 2\pi\alpha'=2\pi l_s^2$
 and Str denotes the completely
symmetrised trace over $N$-dimensional indices.
Amongst the solutions to the equations of motion obtained from
(\ref{dzero}) are those describing $t$-dependent fuzzy 2-spheres.
These are obtained from the ansatz $ \Phi^i = \hat{R}(t)  \alpha_i , i
= 1,2,3 ; \Phi^m = 0,  m = 4..9 $, where $\alpha_i $ are the
generators of $su(2) $ in an $N$-dimensional representation, with
\begin{equation}
 [ \alpha_i , \alpha_j ] = 2 i \epsilon_{ijk} \alpha_k
\end{equation}
 The physical
radius of the fuzzy $S^2$ which we denote by $R(t)$ is related to
$\hat{R} $ via
\begin{equation}\label{phms} 
 R^2 = {\lambda^2 \over N }tr ( \Phi_i^2  ) =
 \lambda^2 C \hat{R}^2 
\end{equation}
 with $ C = ( N^2- 1) $ is the value of the second order Casimir
invariant.

Substituting this ansatz into the action (\ref{dzero}) and using the approximate
relation (valid in the leading large $N$ limit) 
$Str (\alpha_i \alpha_i )^n = N  C^n $, we get
\begin{equation}\label{actzb}
S_0 = - T_0 N 
 \int dt \sqrt { ( 1 -  ( \lambda^2 C ) (\partial_t  \hat R )^2 )~~
( 1 + 4 \lambda^2 C \hat R^4 ) }
\end{equation}
The conserved energy is
$ E = \sqrt{    { \frac{1+ 4 \lambda^2 C  {\hat R}^4}
{1-\lambda^2 C \dot{{\hat R}}^2 } }}
$
and the equations of motion are
\begin{equation}\label{dzeqm}
\frac{d}{dt} \sqrt{    { \frac{1+ 4 \lambda^2 C  {\hat R}^4}
{1-\lambda^2 C \dot{{\hat R}}^2 } }} =0
\end{equation}
Integrating this equation with the initial condition $\hat{R} =
\hat{R}_0 $ and $\dot{\hat{R}} = 0 $ at $t = 0$, we obtain
\begin{equation}\label{deqm}
\dot{R}^2 = 4 \frac{ R_0^4 - R^4 } { {\lambda^2 C + 4 R_0^4 } }
\end{equation}
where we work in physical units. This equation, studied in the
context of time-dependent membranes in \cite{coltuck},  has
solution
\begin{equation}\label{dsol}
             R(t) = R_0 Cn\big( t\sqrt{2}/\tilde{R}_0,
          1/ \sqrt{2} \big)
\end{equation}
where $ Cn(u,k) $ is a Jacobi elliptic function and $\tilde{R}_0^2
=\frac {{ R_0^4 + \lambda^2 C/4 }}{{R_0^2}} $. The initial
conditions are satisfied due to the properties $Cn(0,k) = 1 $ and
$\frac{{d Cn(u,k)}}{ du} = -Sn(u,k) dn(u,k) $ with $Sn(0,k) = 0 $.

The functions $Cn(u,k) $ have the property that they are monotonically
 decreasing with zeros at the special value $ u = K(k)$ where $K $
 is a complete elliptic integral of the first kind. 
 Thus our D2-brane solution describes a spherical
 collapse to a point after a time
\begin{equation}\label{collapse}
               \frac { {\sqrt{2}t_*}}{ {\tilde{R}_0}} = K(1/ \sqrt{2})
\end{equation}
ie the collapse time is $t_* = \frac{1}{\sqrt{2}R_0} \left(
\sqrt{R_0^4 + \lambda^2 C/4} \right)K(\frac{1}{\sqrt{2}})$. Hence
we see that $t_*$ increases as we increase the size of $R_0$
(assuming that $R_0^4 >> \lambda^2 C/4 $). Solutions of a similar 
form have been discussed recently, in the context of
3-form flux backgrounds, in \cite{chenlu}

Now consider the DBI action for D2-branes moving in flat space but
including Abelian world-volume gauge fields -
\begin{equation}\label{dtwo}
S_2 = - T_2 \int d^3\xi \sqrt{ -{\rm det} ( G_{MN} \pt_a X^M \pt_b
X^N + \lambda  F_{ab} )}
\end{equation}
where in (\ref{dtwo}) $ \xi^a , a=0,1,2 $  are worldvolume
coordinates of the D2-brane, $G_{MN} $ is the flat space metric in
$D+1$ dimensions, $ M = 0\dots D$,  $F_{ab}$ is the Abelian gauge
field
 strength and $T_2 $ is the brane tension.
We choose a static gauge where $\xi^0 = t = X^0 , \xi^1 = X^1, \xi^2 = X^2 $. We want to
consider time dependent solutions of a spherical D2-brane and so we shall
choose world volume coordinates $\xi^1 = \theta , \xi^2 = \phi $,
 and embedding $X^1 = R(t) \sin\theta \sin \phi ,
X^2 = R(t) \sin\theta \cos \phi , X^3 = R(t) \cos \phi $, where $R(t) $
is the radius of the
spherical D2, in physical units.

In order to correctly reproduce the dynamics of the time dependent fuzzy spheres
considered elsewhere, we shall take the field strength $F_{ab} $ to define $n$ units
of magnetic flux through the world volume, ie
\begin{equation}\label{flux}
\int d\Sigma^{ab} F_{ab} = 2 \pi n
\end{equation}
where $d\Sigma^{ab} $ is the infinitesimal world-volume surface element.
Since only the $F_{12} $ components are non-zero, $F_{12} =
 \frac{n sin \theta }{ 2  } $. Its easy to see that the action for the
D2-brane reduces to
\begin{equation}\label{dtwob}
          S_2 = - 2 \pi n \lambda T_2 \int dt \sqrt{\Big(1-\dot{R}^2 \Big)
 \Big( 1 + 4\frac{R^4 }{ n^2\lambda^2}  \Big) }
\end{equation}
The equations of motion for this sytem can be written in the form
\begin{equation}\label{motiona}
\frac{d}{ dt}\left(  \frac{\sqrt{ 1+ \frac{4 R^4 }{ n^2\lambda^2
}} }{ {1-\dot{R}^2 }}\right)   = 0
\end{equation}
Comparing the above equation to that describing  a system of $N $
D0-branes, discussed earlier, we see that in the leading $N$
approximation the equations coincide if we identify $N = n $.

It is useful to introduce dimensionless variables $(r, s ) $,
where $ s = {dr \over d T  } $,  such that the Lagrangian is $
\sqrt { ( 1+r^4 ) ( 1 - s^2 ) }$. The variable $T $ is a
re-scaled time $t$. The relations between the dimensionless
variables and the original variables in (\ref{actzb}) are
\begin{eqnarray}\label{dimlsrelms}
r^4 &=& ( 4  \lambda^2 C ) \hat R^4  \nnm\\
s^2 &=& (  \lambda^2 C ) {\dot{\hat R}}^2  \nnm\\
T &=& \sqrt {2} ( \lambda^2 C )^{-1/4} t
\end{eqnarray}
Using the relation between $ \hat R $ and the physical $R$ in
(\ref{phms}) we also see that
\begin{equation}
 s = {d r \over d T } = { d  R \over d t }
\end{equation}
so that the relativistic barrier is just $s=1$.


\section{Corrections from the symmetrised trace}

Consider the ansatz $\Phi_i = \hat{R} \alpha_i$, introduced in the
$D0$-brane theory in the
previous section. Substituting this into the action (\ref{dzero}) yields
\begin{equation}\label{dzeroa}
S_0 = -T_0 \int dt\ Str \sqrt{ (1+ 4 \lambda^2 \alpha_i\alpha_i
{\hat R}^4)( {1-\lambda^2 \alpha_j\alpha_j \dot{{\hat R}}^2) }}
\end{equation}
with both $i$ and $j$ summed from $1$ to $3$. The derivation of
(\ref{dzeroa}) uses properties of the symmetrised trace. For the
analogous discussion in the $D1$-brane case, see \cite{cmtbcore}.
The equations of motion following from (\ref{dzeroa}) are
\begin{equation}\label{dzeqma}
Str\ \frac{d}{dt} \sqrt{    { \frac{1+ 4 \lambda^2
\alpha_i\alpha_i  {\hat R}^4} {1-\lambda^2 \alpha_i\alpha_i
\dot{{\hat R}}^2 } }} =0
\end{equation}

In the large $N$ limit, we may replace $\alpha_i\alpha_i$ by
$C(N)$, and we obtain (\ref{dzeqm}). However, we wish to use more
precise expressions here. To consider this, suppose that one
wishes to evaluate the symmetrised trace of some function
$f(\alpha_i\alpha_i)$ of the summed product $\alpha_i\alpha_i$. If
we expand
\begin{equation}\label{feqn}
 f(x) = \sum_{n=0}^\infty f_nx^n
\end{equation}
for some constants $f_n$, then
\begin{equation}\label{feqn1}
 Str f(\alpha_i\alpha_i) = \sum_{n=0}^\infty f_n Str(\alpha_i\alpha_i)^n
\end{equation}
Generally, one has
\begin{equation}\label{feqn2}
 Str (\alpha_i\alpha_i)^n = N  \sum_{i=1}^{n-1} a_iC^{n-i}
\end{equation}
for some $n$-dependent coefficients $a_i$. In this section, we will explicitly
determine $a_0,a_1$ and $a_2$, whilst in Section 7 we will give
 the full result for
 $Str (\alpha_i\alpha_i)^n$ in the case where the generators $\alpha_i$ are in
 the spin-half representation of $su(2)$.

We will show in the following that
\begin{equation}\label{aeqn}
a_0=1, \quad a_1=-\frac{2}{3}n(n-1), \quad a_2 = \frac{2}{45}n(n-1)(n-2)(7n-1)
\end{equation}
so that
\begin{equation}\label{corr}
  Str ( \alpha_i \alpha_i )^n = N \bigg(
 C^n  - { 2 \over 3 } n ( n -1 ) C^{n-1}
 +   { 2 \over 45 }  n ( n- 1 ) ( n-2 ) ( 7n -1 ) C^{n-2 } +
  \dots \bigg)
\end{equation}
Now note that, if one may write
\begin{equation}\label{Deqn}
 Str ( \alpha_i \alpha_i )^n = D C^n
\end{equation}
for some $n$-independent differential operator $D$, then from
(\ref{feqn1})
\begin{equation}\label{Deqn2}
 Str f( \alpha_i \alpha_i ) = \sum_{n=0}^\infty f_n D C^n = D f(C)
\end{equation}
With the result (\ref{corr}), we see that to the first few orders,
\begin{equation}\label{Deqn3}
D = N \bigg(   1 - { 2 \over 3 } C  { \partial^2 \over \partial C^2 }
        + { 14 \over 45}  C^2  { \partial^4 \over \partial C^4 } +
 { 8\over 9}    C  { \partial^3 \over \partial C^3 } + \dots \bigg)
 \end{equation}
We will use the results (\ref{Deqn2}), (\ref{Deqn3}) in Section
(4.6).

\subsection{Calculation by evaluation on highest weight}

The basic object of study in the symmetrised trace which we are
considering is the element
\begin{equation}\label{defccn}
 \ccn =  { 1 \over (2n)! } \sum  \alpha_{i_1}\alpha_{i_1}
\alpha_{i_2}\alpha_{i_2} \cdots \alpha_{i_n} \alpha_{i_n}
\end{equation}
in the universal enveloping algebra of $su(2)$. The sum is over
the $ (2n)! $ permutations of the $\alpha$'s. Since all the
indices are contracted, this element commutes with all the
generators of $su(2)$, i.e belongs to the centre. By Schur's Lemma
it is proportional to the identity in any irreducible
representation. We can calculate it by obtaining its value on the
highest weight state of a spin $J$ representation. This method of
evaluation on highest weights is quite effective in producing
formulae for the large $J$ limit of $\ccn$, for any $n$, and is
also used in Section 7.

Let us first  set down some conventions and useful facts about
$su(2)$:
\begin{eqnarray}\label{sutwo}
[  \alpha_{i} , \alpha_{j} ] = 2i \epsilon_{ijk} \alpha_{k} \nonumber\\
\alpha_{\pm} = \alpha_{1} \pm i \alpha_{2} \nonumber\\
 ~~[ \alpha_3  ,  {\alpha}_{\pm}  ]  = \pm 2 \alpha_{ \pm } \nonumber\\
 ~~ [ \alpha_{+} , \alpha_{-} ] = 2 \alpha_3    \nonumber\\
 \alpha_{i} \alpha_{i} = (\alpha_{3} )^2 + \alpha_{+} \alpha_{-} +
 \alpha_{-} \alpha_{+}  \nonumber
\end{eqnarray}

The value of $ \alpha_3 $ on the highest weight of the spin $J$
representation is $ 2J$ (where $J$ is $0, \frac{1}{2} ,
 \cdots $) -
\begin{equation}
\alpha_3 | J , J > = 2 J | J , J >
\end{equation}
 The
quadratic Casimir takes the value $ C = 4J ( J +1) $. Evaluation
of $ \ccn $ is  done by writing out the $(2n)!$ permutations and
taking the contracted indices to be equal to $(3,3)$ or $ (+ , - )
$ or $ ( - , + )$. At the end of this process we have a series of
$ \alpha$'s including $ \alpha_3$   or $ \alpha_{\pm}$, an example
of which is
\begin{equation}\label{exampat}
(..) \alpha_+ (..) \alpha_- (..)   \alpha_{+} (..)  \alpha_- (..) |J,J >
\end{equation}
The $(..)$  indicate  powers of $ \alpha_3$. We will be
more explicit later. All the powers of $\alpha_3$ can be commuted
to the left say, by using
\begin{eqnarray}\label{comalphpm}
 \alpha_+ ( \alpha_3 )^I  = ( \alpha_3 - 2 )^I  \alpha_+ \nonumber\\
 \alpha_- ( \alpha_3 )^I = ( \alpha_3 + 2 )^I \alpha_+
\end{eqnarray}

After commuting the $ \alpha_3$ to the left we have an expression
made of a string of $ \alpha_+$ and $ \alpha_-$ operators acting
on the highest weight state. It is useful to calculate
\begin{eqnarray}\label{commutrs}
\alpha_+^m \alpha_-^L |J,J >
= N(L, m )  \alpha_-^{L-m} |J,J >
\end{eqnarray}
where $ N( L, m ) = 2^m { L! \over ( L-m ) ! }
{ (  2J - L + m ) ! \over ( 2J - L ) ! } $.
For example, for the pattern in (\ref{exampat}) we have to evaluate
$  \alpha_+  \alpha_-  \alpha_{+}  \alpha_- | J , J > $
which is equal to $ N(1,1)^2 $.

 For $L=m=k$, the N-factor behaves at large $J$ as $ N(k,k) = (2J )^{k}$.
Consideration of the $N$-factors, together with the explicit
powers of $2J$ from the $ \alpha_3$  shows that the highest power
contributed by a pattern containing $k$ pairs of $ \alpha_{\pm}$
is $ (2J)^{2n-k}$. Hence the $\ccn$ has an expansion of the form $
(2J)^{2n} ( b_0 + b_1 ( 2J)^{-1} + b_2 ( 2J)^{-2} + \cdots ) $ and
we only need to calculate patterns with at most $k$ pairs of $
\alpha_{\pm}$ in order to determine the coefficient $b_k$.

A useful combinatoric factor needed is the number of arrangements
(\ref{defccn}) followed by choices of the values for the $i$
indices, which lead to a fixed $(+-)$  pattern of type
(\ref{exampat}). The factor is $ {k!( 2n - 2k )!  2^k n! \over (n-k)! }$.
Absorbing the $ { 1 \over ( 2n)! }$ from normalization of the
symmetriser we   define
\begin{equation}\label{combfac}
C(k,n) =   { k! ( 2n - 2k )!  2^k n! \over (n-k)! (2n)! }
\end{equation}
The $k$ copies of $ \alpha_+$ can arise from
specifying $ \alpha_i$'s contracted to any of
 $k$ copies of $ \alpha_i$'s which are specified
 to $\alpha_-$, hence the $k!$. The $ 2n-2k$ factors of $\alpha_3$ 
can be contracted with $(2n-2k)! $ possible orderings. 
The first $ \alpha_{\pm}$ contraction can be done with one of the
$n$ pairs of contracted indices and there is a factor of two
because we can assign each index of the pair to a $+$ or a $-$.
Hence we have $2n$ choices  for the first contraction, $ 2(n-1)$
for the second, $ 2(n-2)$ for the third and so on. Collecting
everything we have $ { 1 \over (2n)! } \times
 k! ( 2n-2k)!  \times ( 2n ) ( 2n-2) ( 2n-4 ) ... = { k!  ( 2n-2k)! 2^k n! \over (n-k)! (2n)! }$
as claimed above. Thus (\ref{combfac}) is
 the factor which must multiply the number obtained
by explicit evaluation of the $(+-)$ patterns.

Let us first consider $ k = 0 $. All the terms in the equation
defining $\ccn$ in (\ref{defccn}) are equal to $ ( \alpha_3)^{2n }
= (2J)^{2n } $. $ C(0,n) = 1 $ so   the leading term in the
large $2J$ expansion is just $ (2J)^{2n }$

Now consider $ k =1 $.  Let the $\alpha_-$ sit in the $i_1$'th
position after $ \alpha_3^{i_1-1}$ and let the $ \alpha_+ $ sit in
the $i_2$'th position after an additional $  \alpha_3^{i_2-i_1-1}$
-
\begin{equation}
\sum_{i_1=1}^{2n-1 } \sum_{i_2 = i_1+1 }^{2n}
\alpha_3^{2n-i_2} \alpha_+ \alpha_3^{i_2-i_1-1} \alpha_- \alpha_3^{i_1-1}
|J,J >
\end{equation}
Multiplying by the combinatoric factor in (\ref{combfac}) and
commuting the $ \alpha_3$ to the left we get
\begin{eqnarray}\label{comlft}
&& ~~ C(1,n) \sum_{i_1=1}^{2n-1 } \sum_{i_2 = i_1+1 }^{2n}
\alpha_3^{2n-i_2} ( \alpha_3-2 ) ^{i_2-i_1-1} \alpha_3^{i_1-1}
\alpha_+  \alpha_-  |J,J > \nonumber\\
&& = C(1,n) \sum_{i_1=1}^{2n-1 } \sum_{i_2 = i_1+1 }^{2n}
( 2J)^{2n-i_2} ( 2J -2 ) ^{i_2-i_1-1} (2J)^{i_1-1} N ( 1,1)
|J,J > \nonumber
\end{eqnarray} 
which can be simplified to 
\begin{eqnarray}\label{comlfti}
&&= C(1,n) N(1,1) ( 2J)^{2n-2} \sum_{ \vec i }
( 1 - { 1\over J } )^{i_2-i_1-1}
\nonumber\\
&&= (2J)^{2n-1} \bigl(~~  2n - \frac{8}{3} n ( n-1)   { 1 \over 2J }  +
{ 4 \over 3} n ( n-1)(2n-3)  { 1 \over (2J)^2  }   \nonumber\\
&&~~ - { 16 \over 15 } n ( n-1)( n-2) ( 2n-3)  { 1 \over (2J)^3 } ~~
\bigr )    \end{eqnarray}
The sums here and below can be done with a mathematical
software package such as Maple.
For $ k = 2 $, there are two  patterns :
\begin{eqnarray}
 ( .. ) \alpha_+ ( .. ) \alpha_- ( .. ) \alpha_+ ( .. ) \alpha_- ( .. ) \\
 ( .. ) \alpha_+ ( .. ) \alpha_+ ( .. ) \alpha_- ( .. ) \alpha_- ( .. )
\end{eqnarray}
  We will denote them as the
$ ( +-+-) $ pattern and the $ ( ++-- ) $ pattern.
An alternative notation to distinguish them
is to write
$$
\begin{pmatrix}
 1 & 1 \\
 1 & 1
\end{pmatrix}
$$
 for the first,  and
$$
\begin{pmatrix}
 2 \\
 2  \\
\end{pmatrix}
$$
 for the second.
In the first array,
the first integer in the first column
indicates the number of successive $ \alpha_-$
seen while we read from the right and the second gives
the number of $ \alpha_+$ that follows. The top
integer in the second column gives the number of
 $ \alpha_-$ that follows after this 
 and the lower the number  $\alpha_+$ thereafter.
In the second array, the upper $2$ is the number of $ \alpha_-$
seen reading from the right, and the lower $2$ is the number of
$\alpha_+$ after that. Steps similar to the case $ k=1$ give for
the value of $( ++--)$ on the highest weight
\begin{equation}\label{a21}
N(2,2) C(2,n) ( 2J)^{2n-4} \sum_{ i_1=1}^{2n-3} \sum_{i_2=i_1+1}^{2n-2}
\sum_{i_3=i_2+1}^{2n-1}\sum_{i_4=i_3+1}^{2n} ( 1 - { 1 \over J } )^{i_2+
i_4-i_1 -i_3 -2 }
\end{equation}
Evaluating the pattern $ ( +-+-)$ gives
\begin{eqnarray}\label{a22}
&& N(1,1)^2 C(2,n) ( 2J)^{2n-4} \nonumber\\
&& \times \sum_{ i_1=1}^{2n-3} \sum_{i_2=i_1+1}^{2n-2}
\sum_{i_3=i_2+1}^{2n-1}\sum_{i_4=i_3+1}^{2n}
( 1 - { 1 \over J } )^{i_4+i_2-i_1-i_3-2}  ( 1 - { 2 \over J } )^{i_3-i_2-1}
\end{eqnarray}
These sums can be evaluated and collecting the contributions for
$k=2$ we get
\begin{eqnarray}\label{contrib}
&&( 2J )^{2n-2} \bigl(~ 4n(n-1)  - { 8 \over 3} n(n-1)(4n-7)
 \frac{1}{ (2J)}  \nonumber\\
&& \qquad\qquad +  { 16 \over 45} n(n-1) (n-2) ( 50 n - 101 )
\frac{1}{ (2J)^{2}}
 ~ \bigr)
\end{eqnarray}

For $k=3$ we have $5 $ patterns which we list below together with
the corresponding $N$-factors -
\begin{eqnarray}\label{k3pats}
&&\begin{pmatrix}
 3 \\
  3
\end{pmatrix} ~~~~~~~~~~ N ( 3,3 ) \nonumber\\
&&\begin{pmatrix}
 2 & 1 \\
 2 & 1 \\
\end{pmatrix} ~~~~~~ N ( 2,2 ) N(1,1) \nonumber\\
&&\begin{pmatrix}
2 & 1 \cr
1 & 2 \cr
\end{pmatrix} ~~~~~~ N ( 2,1) N(2,2) \nonumber\\
&&\begin{pmatrix}
1 & 2 \\
1 & 2 \\
\end{pmatrix} ~~~~~~ N( 1,1)N(2,2) \nonumber\\
&&\begin{pmatrix}
1 & 1 & 1 \\
1 & 1 & 1 \\
\end{pmatrix} ~~~ N ( 1,1)^3  \nonumber\\
\end{eqnarray}
Using the compact notation $ \sum_{ \vec i } $
for
\begin{equation}\label{sumsfr3}
 \sum_{i_1=1}^{2n-5}~~\sum_{i_2=i_1+1}^{2n-4}~~\sum_{i_3=i_2+1}^{2n-3}
~~\sum_{i_4=i_3+1}^{2n-2} ~~\sum_{i_5=i_4+1}^{2n-1 } ~~\sum_{i_6=i_5+1}^{2n}
\end{equation}
the result of evaluating the five patterns above is
\begin{equation}\label{respat31}
N(3,3) C(3,n) (2J)^{2n-6}
 \sum_{\vec i} ( 1 - \frac{1}{J} )^{i_2+i_6-i_1-i_5-2}
( 1 -  \frac{2}{J})^{i_3 +i_5 - i_2 - i_4 -2}
 ( 1 -  \frac{3}{J} )^{i_4-i_3 -1}
\end{equation}
\begin{equation}\label{respat32}
N ( 2,2 ) N(1,1) C(3,n) (2J)^{2n-6}
\sum_{\vec i} ( 1 - \frac{1}{J} )^{ i_2+i_4+i_6 - i_1 - i_3 - i_5 -3 }
 ( 1 - \frac{2}{J})^{i_3 - i_2 -1 }
\end{equation}
\begin{equation}\label{respat33}
N( 2,1)N(2,2)C(3,n) (2J)^{2n-6}
\sum_{\vec i} ( 1 - \frac{1}{J} )^{i_2+i_4+i_6 - i_1 - i_3 - i_5 -3 }
 (  1 - \frac{2}{J})^{i_3 +i_5 - i_2 -i_4 -2 }
\end{equation}
\begin{equation}\label{respat34}
 N( 1,1)N(2,2)C(3,n) (2J)^{2n-6}
\sum_{\vec i} ( 1 - \frac{1}{J} )^{ i_2+i_4+i_6 - i_1 - i_3 - i_5 -3 }
 ( 1 - \frac{2}{J})^{i_5  - i_4 -1 }
\end{equation}
\begin{equation}\label{respat35}
N( 1,1)^3 C(3,n) (2J)^{2n-6} \sum_{\vec i}
( 1 - \frac{1}{J} )^{ i_2+i_4+i_6 - i_1 - i_3 - i_5 -3 }
\end{equation}

After expanding in $1/J$ and doing the sums, and collecting
the five contributions for $k=3$ we get
\begin{equation}\label{a3}
(2J)^{2n-3} \bigl(  8 n ( n-1) ( n-2) - 16 ( n ) (n-1)( n-2) ( 2n-5 )
 { 1 \over  2J }   \bigr)
\end{equation}

For $k=4$ there are $14 $ patterns. We write the array description
of the patterns, followed by the corresponding $N$-factors -
\begin{eqnarray}
\begin{pmatrix}
4 \\
4\\
\end{pmatrix} \hskip.2in &&N(4,4)\nnm \\
\bp
3 & 1 \\
1 & 3 \\
\ep \hskip.2in &&N(3,3)N(3,1) \nnm\\
\bp
3&1 \\
3&1 \\
\ep \hskip.2in  &&N(3,3)N(1,1) \nnm \\
\bp
2&2 \\
2&2 \\
\ep \hskip.2in &&N(2,2)^2 \nnm \\
\bp
2 & 1 & 1 \\
2 & 1 & 1 \\
\ep \hskip.2in &&N(1,1)^2 N(2,2) \nnm\\
\bp
2 & 2 \\
1 & 3 \\
\ep
\hskip.2in &&N(3,3) N(2,1) \nnm\\
\bp
2&1&1 \\
1&2&1 \\
\ep
\hskip.2in &&N(2,1)N(2,2)N(1,1)
\end{eqnarray}
and
\begin{eqnarray}
\bp
2&1&1 \\
1&1&2 \\
\ep
\hskip.2in &&N(2,1)^2N(2,2) \nnm\\
\bp
1&3 \\
1&3 \\
\ep
\hskip.2in &&N(1,1)N(3,3) \nnm\\
\bp
1&1&1&1 \\
1&1&1&1 \\
\ep \hskip.2in &&N(1,1)^4 \nnm\\
\bp
1&2&1 \\
1&1&2 \\
\ep \hskip.2in &&N(1,1) N(2,1) N(2,2)\nnm \\
\bp
1&1&2 \\
1&1&2 \\
\ep \hskip.2in &&N(1,1)^2 N(2,2) \nnm\\
\bp
1&2&1 \\
1&2&1 \\
\ep \hskip.2in  &&N(1,1)^2 N(2,2)
\end{eqnarray}

For each of the $14$ terms there is a sum of the type in
(\ref{sumsfr3}) generalised to $8$ summation variables. After
$(2J)^{2n-8} C(4,n)$ multiplied by the appropriate   $ N $-factors
is extracted, what is left is nothing but the number of terms in
the sum which is $ { ( 2n)! \over {( 2n-8)!  8! } } $. Collecting
all patterns relevant to $n=4$, with the appropriate combinatoric
factors, and evaluating on the highest weight gives
\begin{equation}\label{a4}
 16 ( n) (n-1) ( n-2) ( n-3)  ( 2J)^{2n-4}
\end{equation}

With the results in  equations
(\ref{comlfti},\ref{contrib},\ref{a3},\ref{a4}),
 we can find the value of
$ \ccn $ as a function of $J$ to the first few orders
\begin{eqnarray}\label{jexp}
\ccn = (2J)^{2n}  \bigg( 1 &+& (2n) {1\over 2J } + { 4 \over 3 }n(n-1) {
1 \over (2J)^2} -
 { 4 \over 3 }n(n-1){1 \over (2J)^3 }   \nnm\\
&+&  { 16 \over 45 } n(n-1)(n-2)(n+2)  {1\over  (2J)^4 } + \dots\bigg)
\end{eqnarray}
Matching this with an expression of the
form $ a_0 C^{n} + a_1 C^{n-1} + a_2 C^{n-2} + \cdots $
where $ C = 4J(J+1) $ determines
\begin{eqnarray}\label{detas}
a_0 &=& 1 \nnm\\
a_1 &=& -{2 \over 3 } n ( n-1 ) \nnm\\
a_2 &=& { 2 \over 45 } n ( n-1) ( n-2) ( 7n-1 )
\end{eqnarray}

It is worth noting that once we have determined the coefficient
$a_0$  by calculating the patterns with $ k=0$ which determine the
coefficient of $(2J)^{2n}$, the formula in terms of Casimirs fixes
the order $ (2J)^{2n-1}$ term. The latter is checked by
considering patterns with $k=1$. Further consideration of $k=2$
fixes the coefficient $a_1$. With $a_0$ and $a_1$ fixed the term
of order $(2J)^{2n-3}$ is fixed by the expansion in powers 
of the Casimir and  can be checked by
considering patterns with $k=3$. Expanding the contribution from
$k=3$ patterns to next order and collecting the leading order from
$ k=4$ allows the determination of $a_2$. With $a_1,a_2$
determined, we now have a prediction for the coefficient of $
(2J)^{2n-5}$, i.e $ { 16 \over 45 } n(n-1)(n-2) ( n+2) $. We have
checked by considering
 the $ 42$ patterns which arise at $ k=5$ that this
is indeed the correct coefficient. Independent confirmation of
these results using computations based on evaluation of the
Casimirs using chord diagrams are given in the following.


\subsection{Casimirs and chords}

The group invariants discussed above arise in knot theory, in
particular in the study of finite type invariants of knots. These
form a type of basis of knot invariants, and may be understood as
terms from the perturbative expansion of Chern-Simons gauge
theoretic knot invariants. Finite type invariants may be discussed
in terms of chord diagrams \cite{BN}, which are diagrammatic
representations of precisely the individual terms which occur in
$\Str(\alpha_i\alpha_i)^n$. A term in this symmetrised trace will
be the trace of a product containing $n$ pairs of group generators
$\alpha_i\alpha_i$, in some order
$\alpha_{i_1}\alpha_{i_2}....\alpha_{i_{2n}}$. Writing each matrix
generator ${(\alpha_i)}^a_b$ as a vertex
\begin{displaymath}\label{ch1}
{(\alpha_i)}^a_b \quad = \quad
\begin{picture}(2,2)
\put(1,0){\line(0,1){1.5}}
\put(0,-0.5){\makebox(0,0){${}^a$}}
\put(2,-0.5){\makebox(0,0){${}^b$}}
\put(1,1.7){\makebox(0,0){${}^i$}}
\thicklines
\put(0,0){\line(1,0){2}}
\end{picture}
\nonumber
\end{displaymath}
the matrix product joins matrix-labelled legs of the vertices, with the trace
forming a circle.
Chords of the circle are then formed by the pairs of free legs with
like indices. For example,
\begin{displaymath}\label{ch2} \abb
\Tr(\alpha_i\alpha_i\alpha_j\alpha_j)
=\ssk \Picture{\FullCircle\EndChord[3,6]\EndChord[0,9]} \qquad
\Tr(\alpha_i\alpha_j\alpha_i\alpha_j) =\ssk
\Picture{\FullCircle\EndChord[0,6]\EndChord[3,9]}
\end{displaymath}
(conventionally one starts at the point on the circle
corresponding to twelve o'clock and moves counter-clockwise around
the circle as one moves from left to right inside the trace). An
individual term in $\Str(\alpha_i\alpha_i)^n$ will be represented
by a circle containing $n$ chords, ie a chord diagram. Particular
choices of groups and representations then give explicit
realizations of chord diagrams, the resulting assignment of
numerical values to diagrams leading to a \lq\lq weight
system\rq\rq\ and a finite type invariant of knots. In the
following, the \lq\lq order\rq\rq\ of a chord diagram will be the
number of chords, and the \lq\lq value\rq\rq\ $<D>$ of a chord
diagram $D$ will be the
 numerical value
obtained by evaluating the group theoretic trace for
 the group and representation
under consideration (for simplicity, we will
ignore the factor
coming from the trace of the identity matrix
in the expressions
for chord diagrams in this section).

An operator of much interest for knot theoretic applications is that of 
cabling. This is closely
connected with the Adams operation in group theory and with the
fundamental Alexander-Conway invariant in knot theory \cite{KSA}.
Cabling involves taking multiple covers of the encompassing circle
in chord diagrams, and lifting the chord ends in all possible ways
to this new circle. It is clear by definition that the symmetrised
trace $\Str(\alpha_i\alpha_i)^n$ is invariant (up to a factor)
under the cabling operation. In fact it is the unique eigenvector
of
 highest eigenvalue  of the transpose of the cabling matrix \cite{KSA}.
For example, for the chord diagrams with two chords, define the basis

\begin{displaymath}\abb
\begin{pmatrix} 1 & 0
\end{pmatrix}
= \Picture{\FullCircle\EndChord[3,6]\EndChord[0,9]} \qquad
\begin{pmatrix} 0 & 1
\end{pmatrix}
=\Picture{\FullCircle\EndChord[0,6]\EndChord[3,9]} \vspace{24pt}
\end{displaymath}
\noindent
Then the cabling operation $\psi$, which takes the double cover of
the chord diagram
circle and lifts chord ends, is given by
\begin{equation}\label{ch4}
\psi = 4
\begin{pmatrix}
3 & 1 \\ 2  & 2
\end{pmatrix}
\end{equation}
\noindent The eigenvectors of $\psi^T$ are then $(-1,1), (2,1)$,
with eigenvalues $4,16$ respectively. For comparison,
$\Str(\alpha_i\alpha_i)^2 = {1\over3}(2,1)$, which is proportional
to the eigenvector with highest eigenvalue. We note that the
symmetrised trace STr includes a normalization factor $1/(2n)!$.

There are in general relations between chord diagrams \cite{BN},
arising from identities satisfied by the generators $\alpha_i$.
For example, at order three,

\begin{displaymath}
\abb
\Picture{\FullCircle\EndChord[1,11]\EndChord[3,7]\EndChord[5,9]}\quad =
\quad 2\hspace{6pt} \Picture{\FullCircle\EndChord[2,6]\EndChord[4,10]\EndChord[0,8]}
\quad - \quad \Picture{\FullCircle\EndChord[2,8]\EndChord[4,10]\EndChord[0,6]}.
\vspace{24pt}
\end{displaymath}
\noindent Taking the full set of such relations into account, one
can show that the following sets of elements define bases for
diagrams of order $2$, $3$ and $4$ -

\begin{eqnarray*}
&& \left(\ssk
\Picture{\FullCircle\EndChord[3,6]\EndChord[0,9]}\ssk,\quad
\Picture{\FullCircle\EndChord[0,6]\EndChord[3,9]}\ssk
\right)\\[20pt]
&&\left(\ssk
\Picture{\FullCircle\EndChord[1,11]\EndChord[3,9]\EndChord[5,7]}\ssk,\quad
\Picture{\FullCircle\EndChord[2,8]\EndChord[4,10]\EndChord[0,6]}\ssk,\quad
\Picture{\FullCircle\EndChord[2,6]\EndChord[4,10]\EndChord[0,8]}
\ssk\right)\\[20pt]
&&\left(\ssk
\Picture{\FullCircle\EndChord[0,10]\EndChord[1,5]\EndChord[3,7]\EndChord[2,8]}\ssk,\quad
\Picture{\FullCircle\EndChord[0,3]\EndChord[2,5]\EndChord[4,7]\EndChord[6,9]}\ssk,\quad
\Picture{\FullCircle\EndChord[1,7]\EndChord[2,10]\EndChord[4,8]\EndChord[5,11]}\ssk,\quad
\Picture{\FullCircle\EndChord[8,10]\EndChord[3,9]\EndChord[0,4]\EndChord[2,6]}\ssk,\quad
\Picture{\FullCircle\EndChord[2,10]\EndChord[3,9]\EndChord[4,7]\EndChord[5,8]}\ssk,\quad
\Picture{\FullCircle\EndChord[1,11]\EndChord[2,10]\EndChord[4,8]\EndChord[5,7]}\ssk
\right)
\\[20pt]
\end{eqnarray*}
%
We will now specialise to the group $sl(2)$. In this case, it is
possible to derive a reductive formula relating the value of a
chord diagram with $n$ chords to the values of combinations of
chord diagrams with fewer than $n$ chords \cite{CV} (the group
theory conventions of this reference will be used in this section.
This will introduce some factors of $2$ as will be noted below).
Thus one may express the value of any chord diagram as a
polynomial in $c$, where the Casimir $c$ is the value of the
diagram with one chord:
\begin{equation}\label{ch7}
\abb
\vspace{3.5pt}
c = \Picture{\FullCircle\EndChord[3,9]}
\end{equation}
\noindent
Using this reduction to evaluate the diagrams in the
basis elements defined above, one finds the explicit results
\begin{eqnarray*}
&& \hspace{-14pt}\big(c^2,c(c-2)\big) \\ [4pt]
&& \hspace{-14pt}\big(c^3,c(c^2-6c+8),c(c-2)^2\big) \\ [4pt]
&& \hspace{-14pt} \big(c^2(c-2)^2, c(c-2)^3, c(c^3-10c^2+40c-40), c(c-2)^2(c-4),
c^3(c-2), c^4\big)
\end{eqnarray*}
\noindent The expressions for $\Str(\alpha_i\alpha_i)^n$, for $n$
up to $4$, may then be found, either by directly carrying out the
symmetrised trace prescription, or by using the action of the
cabling operator $\psi$ and deducing the highest eigenvalue
eigenvector of $\psi^T$. One finds for general groups
\begin{eqnarray*}
\Str(\alpha_i\alpha_i)^2 &=& {1\over3}(2,1) \\
\Str(\alpha_i\alpha_i)^3 &=& {1\over3}(1,-1,3)  \\
\Str(\alpha_i\alpha_i)^4  &=& {1\over15}(6,2,1,1,3,2)
\end{eqnarray*}
Evaluating these for $sl(2)$ yields
\begin{eqnarray*}
 \Str(\alpha_i\alpha_i)^2\vert_{sl(2)} &=& c^2 -{2\over3}c
\\ [4pt]
 \Str(\alpha_i\alpha_i)^3\vert_{sl(2)} &=& c^3 - 2c^2 + {4\over3}c  \\[4pt]
  \Str(\alpha_i\alpha_i)^4\vert_{sl(2)}  &=&
                             c^4 -4c^3 +{36\over5}c^2-{24\over5}c
\end{eqnarray*}
Evaluating a chord diagram gives a polynomial in $c$.
 The coefficients
of this polynomial are given in terms of quantities defined by chord
intersection properties
of the diagram \cite{CV}. For a chord diagram $D_n$ with $n$ chords,
 one finds
\begin{equation}\label{ch11}
<D_n> = c^n - 2Ic^{n-1} + \Big(2I(I-1) -4T +8Q\Big)c^{n-2} + o(c^{n-3})
\end{equation}
where $I$ is the number of intersections of pairs of chords in the diagram,
 $T$ the
number of triple intersections and $Q$ the number of quartic
intersections (intersections of four chords in the shape of a
square, with possible other intersections amongst these four
chords). This leads to corresponding results for the symmetrised
trace operator. Denote by $<X>_{av(n)}$ the value of some quantity
$X$ defined for chord diagrams, averaged over the set of all chord
diagrams of order $n$ which arise from the symmetrised trace.
Then, for the result quoted above we see that the coefficient
 of $c^{n-1}$
in the polynomial arising from evaluating $\Str(\alpha_i\alpha_i)^n$ is just
$-2<I>_{av(n)}$.

\noindent We have
\begin{equation}\label{ch12}
<I>_{av(n)} = {1\over6}n(n-1)
\end{equation}
To prove this, note that any pair of chords can either intersect or
not in any diagram. Thus, to find $<I>_{av(n)}$, we consider
all possible positions of a pair of chords on a circle where there are
$2n$ possible endpoints for chords. For each such placement of the
pair, one sees if $I$ is $0$ or $1$, and then sums this, and finally one
divides by the total number of such placements.
This leads to the expression for each pair of chords
\begin{equation}\label{ch13}
{\sum_{i=1}^{2n-3} i(2n-2-i)\over (n-1)(2n-1)(2n-3)} = {1\over3}
\end{equation}
This is to be multiplied by the number of choices of pairs of chords,
ie ${1\over2}n(n-1)$, giving the result above.
Alternatively, since the result is proportional to $n(n-1)$, one can find the
coefficient by computation for $n=2$.
This provides a different proof of the result in Section (2.1) above (to compare
coefficients, we note that the Killing form used in \cite{CV} introduces
a factor of $2^{-i}$ multiplying the coefficient of $c^{n-i}$, and so one needs to
divide by this factor in translating to the conventions used elsewhere in
this paper).

For the next order term, involving $c^{n-2}$,
to calculate
$<T>_{av(n)}$ and  $<Q>_{av(n)}$, note that these are proportional to
$n(n-1)(n-2)$ and $n(n-1)(n-2)(n-3)$ respectively. The coefficients are
fixed by explicit calculation for $n=3, 4$ respectively. This gives
the results
\begin{eqnarray*}
<T>_{av(n)}  &=& {1\over90}n(n-1)(n-2)  \\
<Q>_{av(n)}  &=& {1\over15.24}n(n-1)(n-2)(n-3)
\end{eqnarray*}
For $<I^2>_{av(n)}$, note that at large $n$ this behaves as $n^4$, and it
vanishes for $n=0,1$. This fixes all but three of the coefficients of the
expression as a fourth order polynomial in $n$. These further three coefficients
are then found by calculation for $n=2,3,4$. Thus one finds that
\begin{equation}\label{ch15}
<I^2>_{av(n)} = {1\over180}n(n-1)(5n^2-n+12)
\end{equation}

Putting together the above expressions, inserting the
factor of $N$ coming from the trace of the identity,
and using the conventions of the rest of this paper,
we are thus led to the result
\begin{equation}\label{ch16}
{ 1 \over N } \Str(\alpha_i\alpha_i)^n\vert_{su(2)} = C^n -{2\over3}
n(n-1)C^{n-1} + {2\over45}n(n-1)(n-2)(7n-1)C^{n-2}
                             + o(C^{n-3})
\end{equation}
in agreement with the calculations of the previous section.

Clearly the precise form of subsequent terms in this expansion
will depend upon detailed features of diagrams of increasing
complexity. In order to obtain information on the general
structure of these terms, we now outline an argument concerning
the large $n$ behaviour of $\Str(\alpha_i\alpha_i)^n$. First note
that at large $n$, ie for diagrams with a large number of chords,
the symmetrised trace generates sets of diagrams which are
dominated by those with $I$ large and the other quantities $T, Q,$
etc, relatively small, since $I$ is always greater than these
quantities in any given diagram, and there are in addition many
diagrams with $I$ non-zero and the other quantities all zero. To
determine the pure $I$ dependence of a chord diagram, consider the
diagram $P_n$ which corresponds to the expression
\begin{eqnarray}\label{npoly}
\Tr \Big (   (\alpha_{i_1}\alpha_{i_2}\alpha_{i_3}\alpha_{i_1})
         (\alpha_{i_4}\alpha_{i_3}\alpha_{i_5}\alpha_{i_4}) \dots
 \dots (\alpha_{i_{2n-3}}\alpha_{i_{2n-4}}\alpha_{i_{2n-2}}\alpha_{i_{2n-3}})
  && \nonumber \\
 \times(\alpha_{i_{2n-3}}\alpha_{i_{2n-1}}\alpha_{i_{2n-2}}\alpha_{i_{2n}}
      \alpha_{i_{2n-1}}\alpha_{i_{2n}}\alpha_{i_{2}} ) \Big) &&
\end{eqnarray}
This corresponds to a chord diagram, with $n$ chords in the shape
of an $n$-polygon inside the circle, but with one pair of adjacent
chord ends (labelled by $\alpha_2$ and $\alpha_{2n}$ in this case)
interchanged to remove one intersection. The diagram $P_n$ has $I$
non-zero and all other geometric quantities $T,Q$, etc, zero, and
so it can be used to deduce the pure $I$ dependence of a general
diagram. It is not difficult to prove by calculation that the
value of $P_n$ is given by
\begin{equation}\label{npolyeval}
<P_n> = c(c-2)^{n-1}
\end{equation}
This indicates that the pure $I$ dependence of a general diagram $D_n$
with $n$ chords is given by
\begin{eqnarray}\label{Dn}
<D_n> = c^n - 2Ic^{n-1} &+& 2I(I-1)c^{n-2} + \dots +
\begin{pmatrix}
I \\ i
\end{pmatrix} (-2)^ic^{n-i} +
\dots
\nonumber \\ &&
+
\begin{pmatrix}
I \\ n-1
\end{pmatrix} (-2)^{n-1}c
\end{eqnarray}
We now wish to average this over chord diagrams at order $n$.
Following arguments similar to those given earlier, we find that
at large $n$ we have
\begin{equation}\label{Iav}
<I^\lambda>_{av(n)} \simeq (n^2/6)^\lambda
\end{equation}
Using (\ref{Dn}), we can deduce the contributions to the
symmetrised trace from chord diagrams  with up to double
intersections at large $n$. This gives an approximation to the
symmetrised trace which is (in the conventions of the rest of this
paper)
\begin{eqnarray}\label{Strapprox}
{ 1 \over N }  \Str(\alpha_i\alpha_i)^n &\sim& \sum_{i=0}^{n} \frac{1}{i!}
\left(\frac{-2}{3}\right)^in^{2i}C^{n-i} \nonumber \\
 &=& C^n - \frac{2}{3}n^2C^{n-1} +
 \frac{2}{9}n^4C^{n-2} - \frac{4}{81}n^6C^{n-3}+ \dots
\end{eqnarray}
The expression (\ref{Strapprox}) shows that the signs of terms of
decreasing order in $C$ alternate, and that the coefficients fall
to zero somewhat faster than an inverse factorial. This formula
gives a guide to the general structure of higher order terms. From
(\ref{ch16}) the large $n$ limit of the exact expression is $C^n -
(2/3)n^2C^{n-1} + (14/45)n^4C^{n-2} + \dots$, and we see that
ignoring the effects of higher order intersections in diagrams has
reduced the coefficient of the $C^{n-2}$ term by about $25\%$. The
effects of higher order corrections with the structure in
(\ref{Strapprox}) will be reviewed in Section 6.

\bigskip


\section{Energy corrections to first order in $1/N$}

In this section we will study the effects of the corrections to
the theory defined by (\ref{dzeroa}) which are of the lowest order
in $1/N$. The leading order Lagrangian is
\begin{equation}\label{lag}
\cLz  =
- { \sqrt { 1 + 4 \lambda^2 C R^4 }  \sqrt { 1 -  \lambda^2 C \dot R^2 } }
\end{equation}
and the energy to lowest order is
\begin{eqnarray}\label{loworderE}
\cEz &=& \dot R { \partial \cLz \over \partial \dot R }  - \cLz \nnm \\
  &=&  \sqrt  \frac{ 1 + 4 \lambda^2 C R^4 } { 1 -  \lambda^2 C \dot R^2 } 
\nnm\\
 &=& \sqrt{ 1 + r^4 \over 1 - s^2 } 
\end{eqnarray} 
In the last line we used the dimensionless variables in (\ref{dimlsrelms}). 
Including the corrections to next order, by the arguments of
Section 3 we are led to the corrected Lagrangian
\begin{equation}\label{lagc}
\cLo  =  \left( 1 - { 2 \over 3 }  C  { \partial^2 \over \partial C^2 }\right ) \cLz
\end{equation}
We will work with re-scaled variables $ r^4 = 4 \lambda^2 C R^4 $
and $s^2 = \lambda^2 C \dot R^2 $.
The conserved energy to this order is then
\begin{equation}\label{consen}
\cEo    = \dot R { \partial \cLo \over \partial \dot R }  - \cLo
    =  \left( 1 - { 2 \over 3 }  C  { \partial^2 \over \partial C^2 }\right )
 \cEz
\end{equation}
It is useful to use the variables $U$ and $\gamma$, defined by
\begin{equation}\label{rsdefs}
U = \sqrt { 1 +r^4 }, \qquad  \gamma = { 1 \over \sqrt { 1 - s^2
}}
\end{equation}
Then the zeroth order energy is just
\begin{equation}\label{gammaU}
\cEz = \gamma U
\end{equation}
which indicates that one might think of $U$ as a position
dependent \lq\lq mass\rq\rq.

We will denote the first order expression for the energy by the
symbol $E$ in this section. From equation (\ref{consen}) this is
given by
\begin{equation}\label{core}
E ( \gamma , U ) = \gamma U -  { \gamma \over 6 C U^3  } (
\gamma^2 U^2 - 1 ) ( 3 \gamma^2 U^2 - 4 U^2 +1 )
\end{equation}

\subsection{Plots}

It is instructive to study the energy (\ref{core}) as a function
of the variables $(r, s )$, using (\ref{rsdefs}). We will do this
in the following.

\begin{figure}[htbp]
\vspace{1cm}
\centering
\epsfig{file=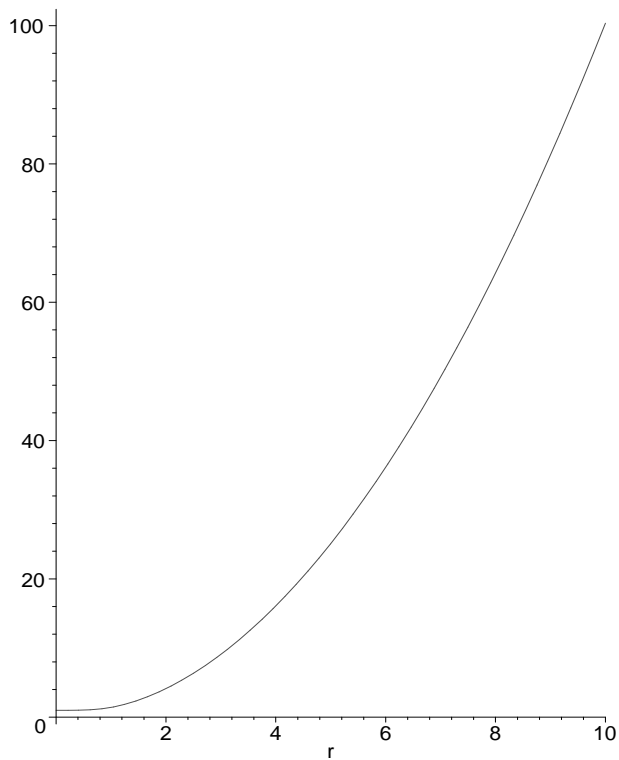, width= 8cm, height = 10cm}
\caption{Plot of the energy as a function of $r$, for $s=0$ and
taking $C = 50$.} \label{fig:plotvsr}
\end{figure}

Plots of the energy function  as a function of $r$ show that at
$s=0$ it is  monotonically increasing. Figure ~\ref{fig:plotvsr}
shows  such a  plot. This behaviour means that as the D2-brane
starts collapsing from rest  at some large radius $r_0$, there is no smaller
radius where the energy can take the same value at $t=0$ while
returning to zero velocity. If the potential had a minimum, a
conventional bounce would be possible with the radial evolution
slowing down to zero velocity and then re-expansion starting with
a change in sign of the radial velocity. So clearly such a bounce
does not happen with the first $1/N$ correction.

\begin{figure}[htbp]
\vspace{1.5cm} \centering \epsfig{file=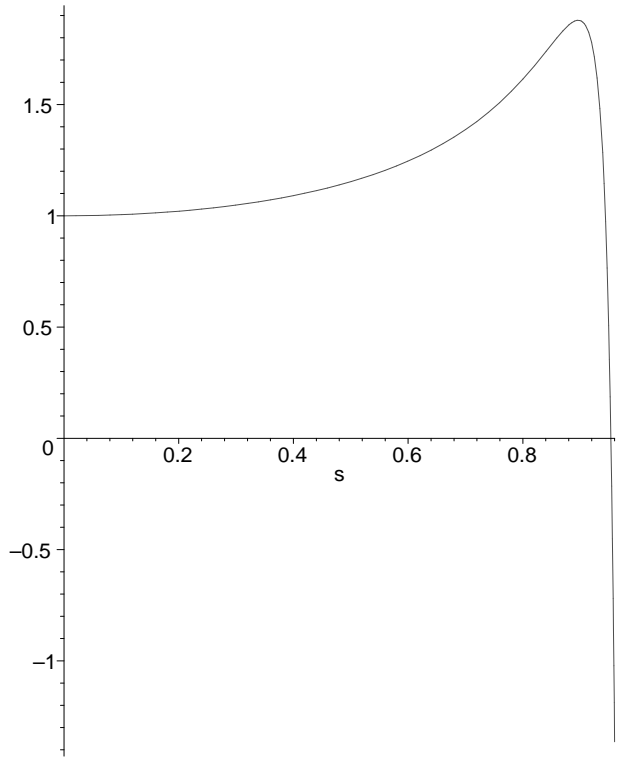,
width=8cm, height=10cm} \caption{Plot of energy as a function of
$s$, for $r=0$ and with $C = 50$.} \label{fig:plotvss}
\end{figure}

Plots of the energy as a function of $s$ at fixed $r$ show an
extremum. Figure ~\ref{fig:plotvss} shows such a plot at $r =0$.
If the energy is bounded above at $r=0$, this means that a
D2-brane, starting from zero velocity at sufficiently large $r_0 $
(hence sufficiently large energy), cannot reach zero radius. This
may seem paradoxical given the conclusions to be drawn from the
plot above of energy as a function of $r$ at $ s =0 $. Further
insight can however be obtained by plotting contours of constant
energy in the $(r,s)$ plane.

\begin{figure}[htbp]
\vspace{2.0cm}
\centering \epsfig{file=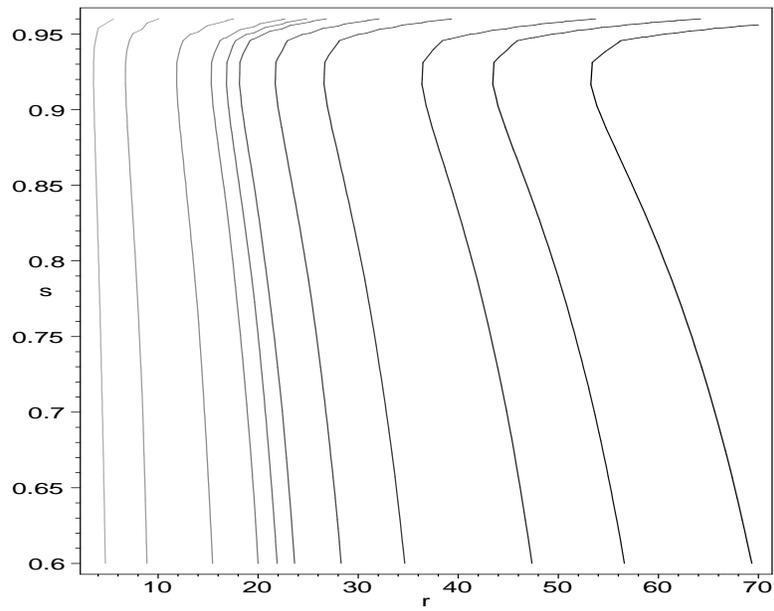, width=10cm, height=8cm}
\caption{Plot of constant energy contours as a function of
$r$ and $s$, with $C =
100$. The energy of the contours is increasing from left to right.}
\label{fig:contplot}
\end{figure}

Figure \ref{fig:contplot} shows contour plots  of the energy
(\ref{core}) as a function of $r$ and $s$. We have chosen $C = 100
$ as an example. If we do not include the $1/C$ correction the
constant energy contours all reach down to zero radius.
The key feature is that $ { \partial E \over \partial s } = 0 $ at
a point where $ s = s_m \ne 0 $. This does not happen for ordinary
mechanical systems described by energy functions of the form $ E =
{ 1 \over 2 } m s^2  + V(r) $. For such energy functions $ {
\partial E \over \partial s }  = 0 $ only at the point $s =0$. As
we will see  from analytic considerations in section 4.2,
solutions to $ { \partial E \over \partial s }  = 0 $ or
equivalently $ { \partial E \over \partial \gamma }  = 0 $ do
exist for $ \gamma \sim  C^{1/4}$.

More detailed numerical investigation of the trajectories are
possible. For fixed initial energy $ E_0$ or initial position
$r_0$ (or equivalently $U_0$) where the velocity is zero, we can
solve the equation $ E ( \gamma , U ) = E_0 $ for $U ( \gamma ,
E_0 )$. The equation is quartic and has four solutions, one of
which is the physical one. This solution can be followed as $
\gamma $ is changed from unity to some large number. For $E_0$
corresponding to $U_0 > 1 $ but 
$ U_0  < \sim C^{1/4}$ we find that
the classical path describes $U$ decreasing monotonically to
$U=1$. These paths, which we may call { \it perturbative paths  }, 
 are small  $ { 1 \over C }$  perturbations of the paths obtained
from $ \cLz$. For larger values of $U_0$, the path encounters a
minimum radius at some finite value $\gamma_m$ and proceeds to
increasing $U$, which in fact approaches infinity at a finite
upper bound.

\begin{figure}[htbp]
\vspace{0.5cm}
\centering \epsfig{file=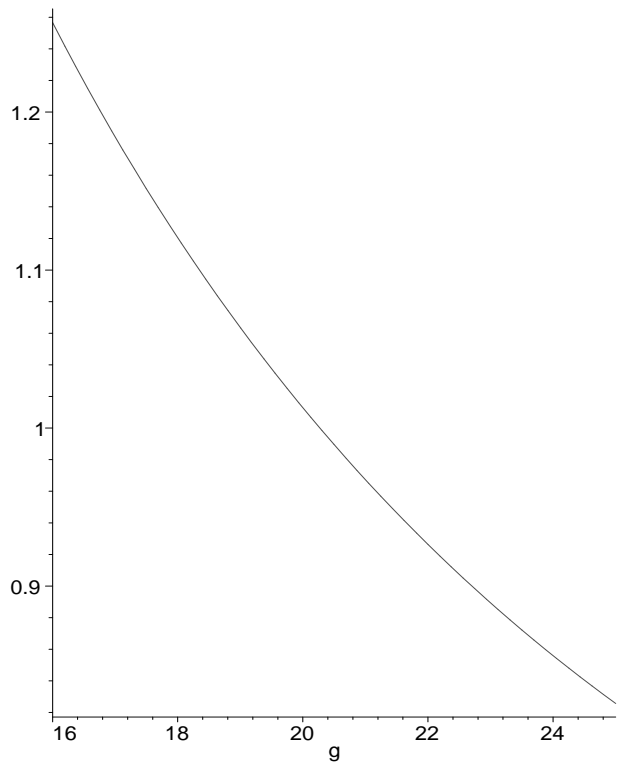, width=8cm, height=10cm}
\caption{Plot of $U$ as a function of $\gamma = g $ along a solution, with $C =50^4$ and
$E_0 = 20$ } \label{fig:contplot2}
\end{figure}

\begin{figure}[htbp]
\vspace{-1cm}
\centering \epsfig{file=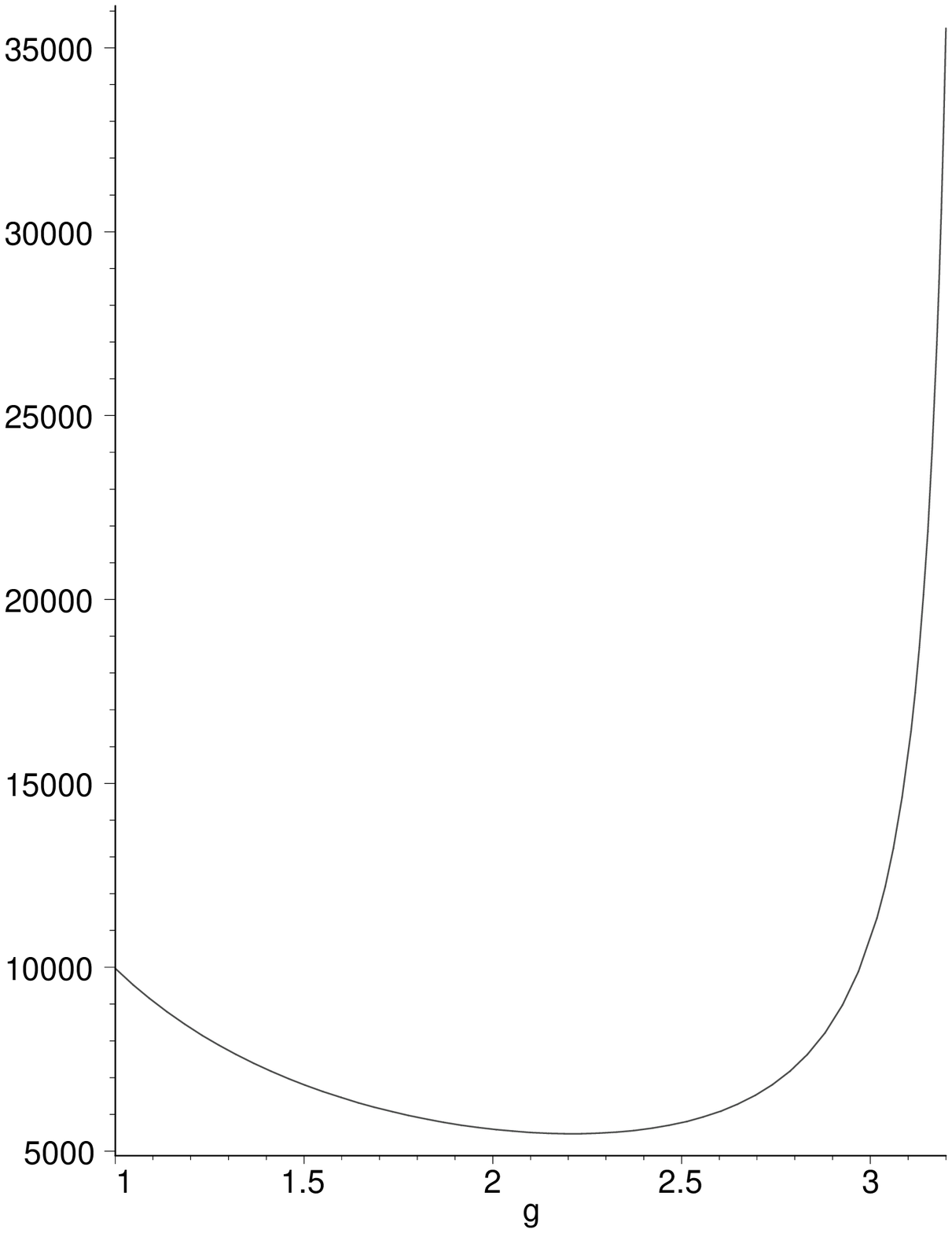, width=8cm, height=10cm}
\caption{Plot of $U$ as a function of $\gamma = g $ along a solution, with $C =50$ and
$E_0 = 100^2  $ } \label{fig:contplot3}
\end{figure}

Figure \ref{fig:contplot2} shows the classical path in the $ (U , \gamma )$ plane
for $ C =50^4 $ and $ E_0 = 20$. For these choices, $U$ decreases
continuously through $U=1$ or $r = 0 $.
Figure \ref{fig:contplot3} shows the classical path in the $ (U , \gamma )$ plane
for $ C =50^4 $ and $ E_0 = 100^2$. For these choices, $U$ decreases
until it reaches a minimum and then increases apparently to infinity
at a finite value of $ \gamma$. Beyond this limiting
$ \gamma $ the quartic equation for $U$ given by
$ E ( \gamma , U )= E_0 $  has no physical solution.
Along these paths we can also compute an effective mass $ \sqrt {
E^2 - p^2 } $ and the (proper) acceleration. We will discuss these
further in Section 8.

\subsection{Extrema}

The extrema in the energy contour are one of the novel features of
$\cLo$. As we will elaborate on later, they occur in a region
where the $1/N$ expansion is becoming problematic, but we will
nevertheless study them in more detail since it is entirely
possible that such extrema occur in the finite $N$ case.

On a contour of constant energy one has $ dE = 0$. To  first
order in variations of $ \gamma , U $ we can write $ dE = ({
\partial E \over \partial \gamma })_U d \gamma + ({ \partial E
\over \partial U })_{\gamma} d U $. At an  extremum of position
along  the energy contour, we have $dU = 0 $. This means that
\begin{equation}\label{van}
({ \partial E \over \partial \gamma })_U = 0
\end{equation}
at that point.
Equivalently
we have
\begin{equation}\label{chn}
 ({\partial U \over \partial \gamma} )_E =
{ ({ - {\partial E} \over {\partial  \gamma} })_U \over ({
{\partial E }\over \partial U  } )_{\gamma}}
\end{equation}
If we consider $U$ as a function of $ \gamma$ along a
constant energy contour, an extremum of
$U$ requires the vanishing of the energy derivative
(\ref{van}).

This gives us a quadratic equation for $U^2$ in terms of $\gamma$
\begin{equation}\label{quadeq}
U^4\Big( -1 + 5 { \gamma^4 \over 2C } - { 2 \gamma^2 \over C }
\Big) + U^2 \Big( { -  \gamma^2 \over C } + {2 \over 3 C } \Big)
-
 { 1 \over 6 C }  = 0
\end{equation}
This is solved by
\begin{equation}\label{solq}
U^2 =  { {  ({ \gamma^2 \over C } - { 2 \over 3 C}  )
\pm \sqrt {  ( { \gamma^2 \over C}  -  { 2 \over 3 C } )^2 +
{ 2 \over {3C} }
( -1 + { 5 \gamma^4 \over {2 C} }
- { 2 \gamma^2 \over C } )       }} \over { 2 ( -1 + 5 { \gamma^4 \over {2C }}
- { 2 \gamma^2 \over C } )} }
\end{equation}

For $ \gamma $ close to $1$ the denominator is negative, while the
argument of the square root is negative as it is dominated by 
$ { -2 \over 3C }$ at large $C$, so there
is no real solution for $U$. The
positive branch must be chosen for a real solution to be possible.
For $C$ large compared to $1$, and $ \gamma^2 \sim \sqrt {C} $,
the denominator changes sign and there is a real solution. The
minimum $ \gamma $ which allows an extremum is obtained by setting
the denominator to zero. This gives
\begin{equation}\label{mingam}
  \gamma^2 = { 1 \over 5 } ( 2 + \sqrt {  4 + 10 C }  )
     \sim \sqrt { 2 \over 5 } \sqrt {  C  }
\end{equation}
where we have used the large $C$ approximation  in the last step.
The associated minimal velocity is $ s^2 = 1 - \sqrt { { 5 \over 2
} } { 1 \over \sqrt { C } }  $.

When $ \gamma $ is very large compared to $C$, we again have no
physical solution, because $U$ becomes close to zero, whereas for 
real $r$  it must be larger than $1$. The upper bound on $
\gamma$ for allowed extrema is obtained by setting $ U=1$ in
(\ref{quadeq}) and solving for $\gamma$. This gives
\begin{equation}\label{gamup}
  \gamma^2   =  { 1 \over 5 } ( 3 +  \sqrt { 4 + 10 C } )
            \sim \sqrt { 2 C \over 5 }
\end{equation}
Comparing the above two equations, we see that there is a very
thin range of $ \gamma $ which allows the extremum to exist.

While there is a finite range of velocities which allows the
extremum to exist, there is no bound on $r$. This is a bit
surprising since we expect the extremum to occur in a sense near
the origin. This can still be true in the following sense. If we
start the collapse with zero velocity at some position $U_0$, and
ask for the minimum $U$ reached, we will get a  minimum which is a
function of $U_0$. Let us denote this minimum as $U_{m} ( U_0 ) $.
We would like to prove that $  { U_m ( U_0)  \over U_0 }  \sim { 1
\over C^{\alpha} } $ where $ \alpha $ is a positive number, so
that this approaches zero in the large $C $ limit.
Now consider a brane starting with zero velocity. Substitute $
\gamma =1 $ in (\ref{core}) to obtain the energy $E_0$  at the
starting position $ U_0$
\begin{equation}\label{enot}
E_0 (  U_0 ) =  U_0 + { 1 \over 6 C  U_0^3 } \Big( U_0^2 -1
\Big)^2 \
\end{equation}
We can determine $ \gamma $ as a function of $U$ along the
curve of extrema by solving the quadratic equation for $ \gamma^2$
in (\ref{quadeq}). This gives
\begin{equation}\label{solgam}
\gamma_m^2 =
{ ( 3 + 6 U_m^2 ) + \sqrt { 24 - 24 U_m^2 + 36 U_m^4 + 90 U_m^2 C }
  \over 15 U_m^2 }
\end{equation}
Substitute this in the energy formula (\ref{core}) to get a
function $ E(U_m) $ given by
\begin{equation}\label{eum}
E ( U_m ) = { { 2(45+90 U_m^2+15 D  )^{1 \over 2 }}
 ( 90 U_m^4 C+18-48 U_m^2+ D  +12 U_m^4+
2 U_m^2 D ) \over 3375U^4 C}
\end{equation}
where $ D $ is given by
\begin{equation}\label{dsq}
D = \sqrt  { ( 24-24 U_m^2+36 U_m^4+ 90 U_m^4 C) }
\end{equation}
Equating $E (U_m ) = E ( U_0 ) $ gives an equation which determines
$ U_m(U_0) $.
This becomes tractable if we use the large $C$ approximation
(assuming $U_m$ is sufficiently small compared to $C$). We find
that $ U_m / U_0 = C^{ -1/4} f_0 ( U_0 ) + C^{ -3/4} f_1 ( U_0 ) +
\cdots $, where $f_0( U_0  ) \sim  1 + \cO  ( { 1 \over U_0^2 } )
$ and $ f_1 ( U_0 ) \sim  1 + \cO ( { 1  \over U_0^2 }   ) $.

\subsection{ Energy-momentum relation }

We now turn to dispersion relations. The momentum is
\begin{eqnarray}\label{mom}
  p &=& { 1 \over \lambda \sqrt C } { \partial L \over \partial s } 
\nonumber \\
  &=&  \gamma U s -  { \gamma s \over 6U^3 C } ( 3 U^2 \gamma^2  +1 )
(  U^2 \gamma^2 -1 )
\end{eqnarray}
We have included a factor of $ { 1 \over \lambda \sqrt C }$ for
convenience. Note that the formula for the momentum, like that for
the energy, has at the leading term the standard form of special
relativity, with $U $ playing the role of mass. We would like to
obtain an equation for $ E(p) $ which can be interpreted as a
deformed dispersion relation.

Let the corrected speed be
\begin{equation}\label{corsp}
 s = { p \over \sqrt { p^2 + U^2  } } + {s_1 \over C }
\end{equation}
where we will determine $s_1$ as a function of $ U $ and $p$. The
corrected $ \gamma $ variable is
\begin{equation}\label{corgm}
 \gamma =  \Big( 1 + { p^2 \over U^2 }\Big )^{1/2} - { p \over U }
\Big( 1 + { p^2 \over U^2 }  \Big) { s_1 \over C }
\end{equation}
Substituting this in (\ref{mom}) and keeping the leading term in
the
 $1/C$ expansion
gives us a linear  equation for $s_1$. The solution of this is
\begin{equation}\label{sone}
s_1 =  { -( p^2 + U^2)^{-3/2}  \over 6 U^2 }      (3U^2+3p^2+1)(U^2+p^2-1)p
\end{equation}
Substituting in the corrected speed (or the corresponding corrected
 $ \gamma $)
in (\ref{core}) we find the first-order corrected energy formula
as a function of $p$ is
\begin{eqnarray}\label{corep}
  E &=& \sqrt { p^2 + U^2 } ~ + ~
 { ( p^2 + U^2 )^{-3/2} \over 6 C U^4 } ( U^2 + p^2 -1 )
( U^2 - 3 p^2 -1 ) ( U^2 + p^2 )^2 \nonumber \\
 & -& { p^2  \over 6 C U^4 } ( 3 U^2 + 3 p^2 +1 ) ( U^2 + p^2 -1 )
\end{eqnarray}

There has been a lot of discussion of deformed dispersion
relations recently in the context of inflation (see
\cite{bmotwo}, for example).
The deformed dispersion relation we are getting
does not have a smooth limit in the zero ``mass'' limit of
 $U = 0$. In our context this limit is of course unphysical
since $ U = \sqrt { 1 + r^4 } $.

\subsection{The time of collapse}

It is also possible to determine the time of collapse in this
system. The leading formulae for velocity $s$ and the
corresponding $ \gamma $ are
\begin{equation}\label{lding}
  s_{(0)} ( U , U_0 ) = {  \sqrt { U_0^2 - U^2 }  \over U_0^2 }, \qquad
 \gamma_{(0)} ( U , U_0 )   = {  U_0 \over U }
\end{equation}
We will find the perturbed formulae
\begin{equation}\label{pertsg}
 s  = s_{(0)} + { 1 \over C } s_{(1)} ( U , U_0 ), \qquad
  \gamma  = \gamma_{(0)} + { s_{(1)} \over C } \gamma_{(0)}^3
( 1 - \gamma_{(0)}^{-2} )^{ 1 \over 2}
  \equiv \gamma_{(0)} + { \gamma_{(1)} \over C }
\end{equation}
where the last line defines $ \gamma_{(1)}$. We substitute
(\ref{enot}) into the left-hand side of (\ref{core}), and on the
right-hand side  put in the $ 1/C$ expansion of the velocity,
keeping only the leading terms. This gives
\begin{equation}\label{corepl}
E_0  ( U_0 ) - U_0  =  { \gamma_{(1)} U \over C } + { 1 \over C }
 f ( \gamma_{(0)} , U )
\end{equation}
where $ f $ is the correction term that appears in (\ref{core}).
This allows us to solve for $ \gamma_{(1)} $, or equivalently $s_{(1)}$.
We find
\begin{equation}\label{solvsone}
s_{(1)} = { ( U_0^2 -1 ) \over 6 U_0^5 } { \sqrt { ( U_0^2 - U^2 )}
 \over   U^2 }
 ( 3 U_0^4 - U_0^2 U^2 + U_0^2 + U^2 )
\end{equation}
Thus the $ 1/C $ corrected formula for the speed is
\begin{equation}\label{oncsp}
{ dr \over dt } = { \sqrt { ( U_0^2 - U^2)} \over U_0 }
 + { 1 \over C } { ( U_0^2 - 1 ) \over 6 U_0^5  } { \sqrt { ( U_0^2 - U^2 )}
 \over   U^2}  ( 3 U_0^4  + U_0^2 +  U^2 ( 1 - U_0^2 )  )
\end{equation}
and the time of collapse is
\begin{eqnarray}\label{time}
\int dt &=& \int dr { U_0 \over \sqrt { U_0^2 - U^2 } } - { 1
\over 6 C }  \int dr { ( U_0^2 -1 ) ( 3 U_0^2 + 1 )
\over U_0 U^2  \sqrt { U_0^2 - U^2 } } \nonumber \\
 && + { 1 \over 6 C } \int dr { ( U_0^2 -1 )^2 \over U_0^3 \sqrt { U_0^2  - U^2 } }
\end{eqnarray}
These integrals can be evaluated in terms of elliptic functions.


\section{Solution in the neighborhood of the extremum}

In this section, we will study the behaviour of the brane in the
region of the radial extremum found above. Generally, consider the
expansion of the energy around the extremum, keeping the second
order terms. Consider a constant energy contour parameterised by
$\lambda$, i.e we think of coordinates $( G  (  \lambda ) , V  (
\lambda ) )$, where $G$ is some function of the $ \dot r $, for
example $ \gamma$ or $ \dot r $. $V$ is some function of $r$, for
example $ U$ or more simply $r$. By expanding the energy in a
Taylor series we find at lowest two orders
\begin{equation}\label{conds}
 d\lambda \Big( { \partial E \over \partial V  } { d V \over d \lambda  }
     + { \pt E  \over \pt  G } { d  G \over d \lambda  } \Big) = 0
\end{equation}
and
\begin{equation}\label{condsii}
 d \lambda^2 \Biggl ( \frac{1}{2}
 { \pt E \over \pt V } { d^2 V \over d \lambda^2 } +
\frac{1}{2} { \pt E \over \pt G } {d^2 G \over d \lambda^2} + { 1
\over 2 } { \Big( {d V \over d \lambda } \Big)^2} { \pt^2 E \over
\pt V^2 } + { 1 \over 2 } \Big( {  d G \over d  \lambda } \Big)^2
{ \pt^2 E \over \pt G^2 } +  { \pt^2 E \over \pt G \pt V } { d G
\over d \lambda }
 { d V  \over d \lambda  }    \Biggr ) = 0
\end{equation}
Specializing to a point where $ { dV \over d \lambda } = 0 $, then
from (\ref{conds}) we have $ { \pt E \over \pt G  } = 0 $. Then we
obtain
\begin{equation}\label{condsiii}
 \frac{1}{2} { \pt E \over \pt V } { d^2 V \over d \lambda^2 }
+ { 1 \over 2 } { \Big( {d V \over d \lambda }\Big)^2} { \pt^2 E
\over \pt V^2 } + { 1 \over 2 } \Big( {  d G \over d  \lambda }
\Big)^2 { \pt^2 E \over \pt G^2 } +  { \pt^2 E \over \pt G \pt V }
{ d G \over d \lambda }
 { d V \over d \lambda  }  = 0
\end{equation}
If we approximate the equation  in the neighborhood of
the extremum where ${  dV \over d \lambda }  = 0$
by setting this first derivative to zero we get
\begin{equation}\label{loceq}
 { d^2 V \over d \lambda^2 } = -
 { { \pt^2 E \over \pt G^2 } ( {  d G \over d \lambda } )^2
  \over  {  { \pt E \over \pt V }}  }
\end{equation}
The complete equation derived at order $d \lambda^2$ is really
(\ref{condsiii}) but  we may hope that some qualitative properties
such as the existence of two branches corresponding to reflection
with increasing speed or reflection with decreasing speed can be
captured by the simpler equation (\ref{loceq}). This equation
should allow us to prove that the extremum is a minimum of $U$ and
that we need a change of branch of the solution at the extremum.

Let us choose $ V = r $ , $ G = s^2 \equiv S $ and  $\lambda=S$ so
that (\ref{loceq}) becomes
\begin{equation}\label{loceqS}
{ d^2 r \over dS^2 } = - 2\alpha
\end{equation}
where  $\alpha = \frac{1}{2} { \pt^2 E \over \pt S^2 }/ { \pt E \over \pt r }  $
evaluated at the extremum. Equation (\ref{loceqS}) can be solved
in the neighborhood of the minimum to give
\begin{equation}\label{solneighb}
( r - r_m ) = - \alpha ( s^2 - s_m^2 )^2
\end{equation}
We know that $ r > r_m $ because plots show that the constant
energy contours have a  minimum and $\alpha$ is less than zero.
We can
now write the squared velocity
\begin{equation}\label{veleq}
s^2 = s_m^2 \pm \sqrt { { r_m - r \over \alpha }  }
\end{equation}
Hence
\begin{equation}\label{drdt}
 { dr \over dt } = \pm \sqrt {   s_m^2 \pm
\sqrt { r_m - r \over  \alpha }  }
\end{equation}
which can be solved for the time elapsed
\begin{equation}\label{tm}
\int dt = \pm \int dr   { 1 \over    \sqrt {   s_m^2 \pm
\sqrt { r_m - r \over \alpha  }  }
}
\end{equation}
When we choose the outer  negative sign, $r$ decreases with
increasing time, whereas when we choose the outer positive sign,
$r$ increases with increasing time. The collapsing solution cannot
be continued past $r_m$ because the formula for time elapsed would
become complex. These two solutions can be patched at $r= r_m$.
The radial velocity is discontinuous at the patching, but the time
evolution described by the patched solutions is consistent with
energy conservation.

Note that there is also a choice of sign inside the square root.
If we switch the choice of this sign in patching solutions at the
extremum, then the collapsing brane re-expands along the second
branch, which means that, when it reaches the original starting
position (where it had zero velocity), it is travelling close to
the speed of light. If we do not make this switch of sign inside
the square root upon patching, the brane re-expands along the same
branch as it was collapsing and reaches zero velocity at the
starting position.

If we use the form $ ( V , G )  = ( s , r ) $ and use $\lambda = s
$, the differential equation (\ref{loceq}) becomes
\begin{equation}\label{loceqs}
{d^2 r \over ds^2} = - 2\beta
\end{equation}
where $\beta = \frac{1}{2}{ \partial^2 E \over \partial s^2 } / { \partial E
\over \partial r}  $. Then
\begin{equation}\label{chssr}
\int dt = \int dr { 1 \over {  s_m \pm { \sqrt {  r - r_m  }  \over \beta }  } }
\end{equation}
Here the overall choice of sign visible in (\ref{tm}) is not
apparent. However we know by the time reversal invariance of the
Lagrangian that solutions which have an extra $\pm$ in front of
the integral are also allowed. This time reversal invariance was
kept manifest when we worked with the variables $ (r,s^2 )$.

A more precise treatment of the local differential equation in the
neighbourhood of the extremum continues to reveal that $s-s_m$ is
double valued there. With the choice $ ( G , V ) =  ( s, r )$, $
\lambda = s $, if we keep terms involving the first derivative we
get an equation of the form
\begin{equation}\label{fstder}
a  { \partial^2 r \over \partial s^2 } + b \Big( { \partial r
\over
\partial s } \Big)^2 + c { \partial r \over \partial s } +  d = 0
\end{equation}
where the constants are given by
\begin{equation}\label{cnstls}
 a = { E_{r}^{(m)} \over 2 }, ~~~~
 b =  { E_{rr}^{(m) } \over 2 },  ~~~~
 c = E^{(m)}_{rs},  ~~~~~
 d = { E^{(m)}_{ss} \over 2 }
\end{equation}

Equation (\ref{fstder}) is the Riccati equation with constant
coefficients. We can solve this as follows. Use variables ${  y =
{dr \over ds} } $ to rewrite (\ref{fstder}) as
\begin{equation}\label{fstderi}
a { dy\over ds } + by^2 + c y + d = 0
\end{equation}
The roots of the quadratic polynomial in
$y$ play an important role in the solutions.
These roots are
\begin{equation}\label{rts}
a_+ = { - c + D \over 2 b }, \qquad
a_- = { -c - D \over 2  b }
\end{equation}
where we defined $ D = \sqrt { c^2 - 4 db } $.
Numerical check shows that $ D $ is real.
Integrating once gives
\begin{equation}\label{fstint}
\int ds = -{ a \over b } { 1 \over ( a_+ - a_- ) }
 \ln  \left( {  y - a_+ \over y - a_- } \right)
\end{equation}
Imposing the condition that $y = { dr \over ds } = 0$ at   $ s =
s_m $ fixes the integration constant to give
\begin{equation}\label{fstinti}
( s - s_m )  = \frac{a}{D} ~~ \left( ~ \ln ~\Big( { a_+ \over a_-
} \Big) -
  ~\ln~ \Big({  y - a_+ \over y - a_- } \Big)~ \right)
\end{equation}
One more integration and the condition that
$ r = r_m $ at $ s = s_m$ gives
\begin{eqnarray}\label{solfi}
r - r_m = a_+ ( s - s_m) &+&{ a \over b}~ \ln \left(~ 1 -
\frac{a_+}{a_-}
\exp\Big({ \frac{b ( a_- - a_+ ) }{a} ( s -s_m )\Big) } ~  \right) \nonumber\\
 &-&  \frac{a}{b}~
\ln \left( ~ 1 - { a_+ \over a_-}~ \right)
\end{eqnarray}
The original differential equation (\ref{fstderi}) is symmetric
under exchange of the roots $ a_{\pm}$ and as expected the final
solution can be manipulated into the form
\begin{eqnarray}\label{solfii}
r - r_m = a_- ( s - s_m) &+& { a \over b} \ln \left( 1 -
\frac{a_-}{a_+}
\exp \bigg({ \frac{b ( a_+ - a_- ) }{a} ( s -s_m )\bigg) } \right) \nonumber\\
&-&  \frac{a}{b} ~ \ln \left(~ 1 - { a_- \over a_+} ~ \right)
\end{eqnarray}
Expanding the right-hand side of (\ref{solfi}) or (\ref{solfii})
gives
\begin{equation}\label{solexp}
( r - r_m ) = - {E_{ss}^{(m)} \over 2 E_{r}^{(m)}}   \Big( s -
s_m\Big)^2
\end{equation}
in agreement with the approximation which dropped the ${ dr \over
ds } $ terms in the equation. From (\ref{solexp}) it is clear that
for a fixed $r$ there are two values of $s$, one above $s_m$ and
one below.

We can also use the time $t$ as the parameter $ \lambda $ and
obtain an equation which looks a little more complicated but can
nevertheless be solved. It seems that  the time variable may be
less useful since $ dr/dt $ has a discontinuity at the extremum.
In any case the equation is now
\begin{equation}\label{tomederv}
E_{r}^{(m)} \ddot { r } + { 1 \over 2 }  E_{rr}^{(m)}( \dot r )^2
+ { 1 \over 2 }   E_{ss}^{(m)}  ( {\ddot r} )^2 + E_{rs}^{(m)} \ddot r \dot r
=0
\end{equation}
This can again be solved.
 
\subsection{ Quantum mechanics near the extremum }

Equation (\ref{loceqs}) can be viewed as describing the
evolution of $r$ as a function of a ``time'' variable $s$. We can
write down a Lagrangian which leads to the equation and a
corresponding Hamiltonian which generates translations along the
$s$ direction
\begin{equation}
{ d^2 \psi \over d r^2 } + b r \psi  = E \psi
\end{equation}
The solutions to this equation are Airy functions. The correct
boundary condition is to require that the solutions die off for $
r < r_m$. This implies oscillatory behaviour with a damped
amplitude as the magnitude of $s-s_m$ increases. It is important
to note that in this quantum mechanical set-up $s < s_m$ and $ s >
s_m$ are treated symmetrically.

When the terms $ { dr \over ds}  $ are kept the system becomes
dissipative. Such systems cannot be quantised in an ordinary way,
but they can be described quantum mechanically by enlarging the
system to include a bath of bosonic oscillators, see for example
\cite{cl,dissipref}. (  A connection between string theory 
and this quantization of dissipative systems has in fact been made 
in \cite{calthor} ).   
 The correlation
functions $ < r(s) r(s^{\prime } > $ can be considered. Such
treatments in terms of a bath of oscillators can be viewed as
simulating the interaction of the massless degrees of freedom
including $r$ with the stringy spectrum of open string modes. It
is possible that the extra degrees of freedom required are just
the additional modes that live on the brane world-volume, e.g the
various fluctuations of the fields (which have been neglected in
the solutions where we study purely radial evolution).
 The important conclusion for our purposes
is that whereas the classical evolution is ambiguous at the
turning point, a quantum mechanical set-up allows computations of
correlators associated with different points along the trajectory
on either side of the extremum. This leads us to anticipate  that
a more complete stringy  treatment will allow the formulation of
probabilities for the evolution along the two branches.

It may appear that the conclusions from attempting the above
quantum mechanical discussion are artefacts of the exotic choice
of $s$ as time. However, the ordinary time can also be used to
describe the equations, which are then of higher order. This again
suggests that we need additional degrees of freedom. In fact it
has been argued that particle systems with higher derivative
actions share some stringy properties such as the Virasoro
algebra \cite{hovir}.
 The conclusion from this discussion is again that an
attempt to do quantum mechanics leads to the need for extra
degrees of freedom, which are indeed expected in this stringy
context. While these stringy modes are not relevant for most of
the trajectory, at the extremum the only possible classical
evolutions involve discontinuities in velocity, hence formally
infinite accelerations. Since small accelerations are a
requirement for the low energy effective action to be useful, in
this case we expect that extra degrees of freedom become relevant.


\section{Higher order corrections to the Lagrangian}

Based on the results of Section 3, the next order corrections to
the Lagrangian of (\ref{lagc}) give the result
\begin{equation}\label{lagcc}
\cLt =  \left( 1 - { 2 \over 3 }  C  { \partial^2 \over \partial C^2
}    + { 14 \over 45}  C^2  { \partial^4 \over \partial C^4 } +
 { 8\over 9}    C  { \partial^3 \over \partial C^3 }
 \right ) \cLz
\end{equation}
The energy function to this order is
\begin{eqnarray}
\cEt &= & \gamma U - { \gamma \over 6CU^3}(\gamma U+1)(\gamma U-1)
(3\gamma^2U^2+1-4U^2)      \nnm\\
    \quad &+& { \gamma \over 120C^2U^7} (\gamma  U-1)(\gamma U+1)
    \Big(245\gamma^6U^6-640
    \gamma^4U^6
+105\gamma^4 U^4 \nnm \\
&+& 528 \gamma^2 U^6-256 \gamma^2 U^4+63 \gamma^2 U^2-128U^6+ 176
U^4-128 U^2+35\Big) \nnm
\end{eqnarray}
 It is possible to repeat the analysis of Section 4,
studying the effects of the above terms of order $1/C^2$. We omit
the details. In summary, as in the study of the Lagrangian $
\cLo$, there is a class of { \it perturbative paths } in $ ( \gamma, U )$
space (for $ U_0 < \sim C^{1/4}$), which are
 similar to the ones arising from the Lagrangian $ \cLz $.
For higher $U_0$ the paths still end at $U=1$ without going
through the extremum seen with $ \cLo$, but there are important
differences. The proper acceleration does not increase far beyond
$1$ and the final $\gamma$  at $U=1 $  increases with $U_0$
much slower than would be anticipated by extrapolation 
from the final $ \gamma $ for the perturbative paths. This suggests 
some sort of repulsive force. It also indicates that there 
 may be a class of non-perturbative
 (in the sense of the $1/N$ expansion)
paths having small proper accelerations which allow simple reliable 
treatment with the low energy effective actions  neglecting the 
higher derivative terms. 

Going to one further order, using the term suggested by the
expression (\ref{Strapprox}), however restores the extremum when
these terms contribute. Thus, as this emphasises, in regimes where
higher order terms are not negligible it is necessary to obtain
information about the {\it exact} structure of
$Str(\alpha_i\alpha_i)^n$. To this end, in  Section 7 we will
derive and study the exact expression for
this when the spin half representation is
used.


\section{Exact evaluation of the symmetrised trace for spin half}

In this section, the $\ccn $ operator will be evaluated exactly
for the spin $1/2 $ case. We will be using some of the
notation and techniques from Section (3.1). Consider patterns of the
form
\begin{equation}\label{kpat}
\alpha_3^{2n-i_{2k}} \alpha_+ \alpha_3^{i_{2k}-i_{2k-1}-1 } \cdots
\alpha_3^{i_3-i_2-1}\alpha_+ \alpha_3^{i_2-i_1-1}\alpha_- \alpha_3^{i_1-1}
\end{equation}
These are the only patterns that contribute for spin $1/2$ since
any pattern containing $ (..) \alpha_- (..)  \alpha_- (..)$ (where
$(..)$ as in Section 2 stands for arbitrary powers of
 $ \alpha_3 $) gives zero when acting on the highest weight
of the two-dimensional representation. This is an important
simplification compared to the case of general $J$.
Expressions of the form (\ref{kpat})
 are to be summed with summation symbol $ \sum_{ \vec i } $
defined as
\begin{equation}
\sum_{i_{2k} = 2k}^{2n} \sum_{i_{2k-1} = 2k-1}^{i_{2k}-1} \cdots
\sum_{i_2=2}^{i_3-1}   \sum_{i_1 =1}^{i_2 -1}
\end{equation}
\noindent After commuting all the $ \alpha_3$'s to the left
we get
\begin{eqnarray}\label{nxtstp}
&& \sum_{\vec i } ( \alpha_3 - 2 )^{ ( i_2-i_1-1) + (i_4-i_3-1) + \cdots +
( i_{2k} - i_{2k-1} -1 ) } \alpha_3^{ i_1 + ( i_2-i_1-1) + \cdots +
(i_{2k}-i_{2k-1}-1)  } \nnm\\
&& \qquad\qquad \qquad
 \times \, \alpha_+\alpha_-\alpha_+\alpha_- \cdots \alpha_+\alpha_-
\end{eqnarray}
Further simplifications are $ \alpha_3 = 2J =1 $ and $ \alpha_3 -2
= (-1) $. As a result the expression in (\ref{nxtstp}) evaluates
on the highest weight state to
\begin{equation}\label{evsphf}
 2^{k} \sum_{\vec i }\Big( -1 \Big)^{ -k + \sum_{p=1}^{2k} i_p  }  \nnm
\end{equation}
We also used $ N(1,1) = 2 $ for $ J ={ 1 \over 2 } $. The sum over
$i$ can be simplified by defining variables
\begin{equation}
 \hat i_s = i_s -s, \quad s=1,...,2k
\end{equation}
With these variables the expression in (\ref{evsphf})
simplifies to
\begin{equation}\label{respatsphf}
 2^{k} \sum_{\hat i_{2k}}^{2n- 2k } 
\cdots \sum_{\hat i_2 =  0 }^{\hat i_3 } \sum_{\hat i_1 = 0 }^{\hat i_2}
 (-1)^{ \hat i_1 + \hat i_2 + \cdots
\hat i_{2k} } \nnm
 =  { 2^k n! \over (n-k)! k! }
\end{equation}

In relating the patterns to the symmetrised trace there is a
combinatoric factor  (\ref{combfac}). Taking that into account
along with the above, we have for spin $1/2$ that the symmetrised
power of the quadratic Casimir $\ccn$ is \footnote{ Previous versions of this 
paper had a missing $(2n-2k)!$ on the left resulting in a more complicated 
$f(n)$. The corrected analysis yields simpler energy functions, with the 
same general features.}  
\begin{equation}\label{fn}
\ccn = { (n!)^2 \over (2n)! }  \sum_{k=0}^n { 2^{2k} ( 2 n - 2k ) ! 
 \over (( n-k
)!)^2 }  =  ( 2 n + 1 ) 
\equiv f(n)
\end{equation}

Now we can calculate  the Lagrangian. Let $ \hat C = \alpha_i \alpha_i $.
Then
\begin{eqnarray}
L  &=& -  \Str \sqrt { 1 + \hat C  r^4 } \sqrt{ 1 - \hat C s^2 } \nnm\\
&=& - \Str \sum_{l=0}^{\infty} \sum_{m=0}^{\infty} r^{4l} s^{2m}
 (\hat C)^{l+m}      {  1/2 \choose l } {  1/2 \choose m  }  (-1)^m \nnm\\
&=& - N  \sum_{l=0}^{\infty} \sum_{m=0}^{\infty} r^{4l} s^{2m}
      f(l+m)      {  1/2 \choose l } {  1/2 \choose m  }  (-1)^m \nnm\\
&=&  -  N \sum_{n=0}^{\infty} \sum_{ l=0}^{n} f(n) r^{4l}
 s^{2n-2l}  {  1/2 \choose l } {  1/2 \choose n-l }  (-1)^{n-l}
\end{eqnarray}
The factor of $N (=2)$ comes from the trace, and the binomial
factors are expressed in terms of Gamma functions as
\begin{equation}
{  1/2 \choose l } =  { \Gamma ( 3/2 ) \over \Gamma( 3/2 - l )
\Gamma ( l+1 ) }
\end{equation}
The energy can be computed to be
\begin{eqnarray}
E(r,s) &=& - N  \sum_{l=0}^{\infty} \sum_{m=0}^{\infty}
 f (l+m)  ( 2m -1)  {  1/2 \choose l }{  1/2 \choose n-l }
    ( -1)^{m} r^{4l} s^{2m} \nnm\\
&=& - N  \sum_{n=0}^\infty \sum_{l=0}^{n} f (n)  ( 2n-2l-1)  { 1/2
\choose l }{  1/2 \choose n-l }
    ( -1)^{n-l} r^{4l} s^{2n-2l} \nonumber \\
\end{eqnarray}
If we sum over powers of $r $ first and then over powers 
of $s$ we get 
\begin{equation}
E ( r , s ) =   2  { ( 1 + 2 r^4 ) \over \sqrt {( 1 + r^4 )} } 
  {}_{2}F_{1} ( { 1 \over 2 } , { 3 + 4 r^4 \over 2 + 2 r^4 } ;
{ ( 1 + 2 r^4 ) \over ( 2 + 2 r^4 ) } , s^2 )
\label{rfst}
\end{equation}
If we sum over powers of $s$ first and then over powers of $r$ 
we get 
\begin{equation} 
E ( r , s ) =  2  { 1 \over  (  1 - s^2 )^{3/2} }
   {}_{2}F_{1} (  { 2 s^2 - 3 \over 2s^2 - 2 },
 { - 1 \over 2 }  ;
{  -1 \over   2 s^2  - 2  } , - r^4  )
\label{sfst}
\end{equation}  
These are both useful expressions.

It is useful to recall the following facts about the hypergeometric function
$ {}_2F_{1}(z) $.
This function has branch points at $z=1$ and $z=\infty$, and there
is a branch cut extending from $ 1 $ to $\infty$. Formulae are
available for the discontinuity across the branch cut. In
(\ref{rfst})
 attempting to continue past  $s=1$ runs into the branch cut. Continuation
past $ s =1 $ to $s>1$ requires specification of the sheet.
Physically we are not interested in these super-luminal speeds in
any case. In (\ref{sfst}) the hypergeometric function can be
continued past $r=1$ to large positive $r$ even though the the
original sum over powers of $r$ diverges for $ r > 1 $.

It is instructive to specialize (\ref{rfst}) or (\ref{sfst})
to the case of $r  = 0$ or $ s = 0 $. 
\begin{eqnarray} 
E ( r = 0 , s ) &=& 2  { 1 \over ( 1 - s^2 )^{3/2}  } \label{Es}\\
E ( r , s = 0 ) &=& 2 { ( 1 + 2 r^4 )  \over \sqrt { 1 + r^4 }} \label{Er}   
\end{eqnarray} 

From (\ref{Er}) we see that the energy grows monotonically with $r$. 
This means there is no smooth bounce. 
From (\ref{Es}) there is no extremum  as a function of $s$ at $r=0$. 
Hence there is no exotic bounce at $r=0$. 
Further numerical investigation shows that there is no such 
extremum when $ r > 0 $. Hence the exotic bounces of the kind studied 
in section 4  with the first $1/N$ correction, do not occur for 
spin $1/2$. Whether such extrema occur
for other finite  values of $N$ is an interesting open question.

The appearance of the limiting velocity $s=1$ 
in (\ref{rfst}) -- (\ref{Es}) shows that 
the symmetrised trace prescription correctly captures the 
expected relativistic barrier. This does
not seem evident from the original form of Lagrangians such as
(\ref{dzero}).
%


\section{Regimes of validity }

Neglecting higher spatial derivatives implies that the
D2-brane picture only works for distances larger than the
string scale, or $ r > 1 $ in dimensionless units.
We also need to consider higher orders in time derivatives.

We first discuss issues of regimes which are related to radial
velocities and their comparison to $C$, before considering
accelerations. For the original unperturbed solution, described in
Section 2, $E = \sqrt { ( 1 + r^4 ) \over ( 1 - \dot r^2 ) } $ and
if $r_0$ is the initial position where $ \dot r = 0 $, then $ \dot
r^2 =  { r_0^4 - r^4 \over 1 + r_0^4 } $.
 The subsequent analysis
of $1/N$ corrections has shown that higher orders in the $1/N$
expansion are important when $  \gamma^2 \sim  \sqrt{C} \sim N  $.
This gives a lower bound on $r $ for the large $N$ zero-brane
description to remain valid -
\begin{equation}
r^4 > { r_0^4 \over N }  - 1
\end{equation}
This lower bound can be much less than $1$, so we can follow the
evolution for distances much shorter than the string scale.
However it appears that we can follow it all the way to zero if
$r_0^4 $ scales like a power of $N$ which is smaller than one.

Indeed, for such choices of $r_0$, the paths which solve the
equations of motion coming from $ \cLo $ and $\cLt $ have a
similar behaviour to the leading order paths. They describe
2-branes collapsing to zero size with maximal proper accelerations
which grow as $r_0$ grows. The maximal speed reached at zero
radius is very close to the speed of light, with $ \gamma \sim
r_0^2 $. However when $r_0$ is larger than $\sim N^{1/4}$, the
paths described show qualitative changes. Whether we work with $
\cLo $ or $ \cLt $, we find that there is an upper bound on $
\gamma $ which is much less than $ r_0^2 $.

\subsection{ Proper  Acceleration and effective mass }

Two useful quantities to study are the proper acceleration
and the effective squared mass.
The proper acceleration  $ \alpha$ is given
by $ \alpha^2 = { d^2 x_{\mu} \over d \tau^2} { d^2 x^{\mu} \over d \tau^2 }$.
Useful expressions for $ \alpha $ are
\begin{equation}\label{alphus}
\alpha = \gamma^3 { d^2 r \over d t^2}
       =  \gamma^3 s { ds \over dr}
       =  - \gamma^3  s  \frac{E_r}{E_s}
       =  - \frac{E_r}{E_\gamma}
\end{equation}
This is the relativistically invariant acceleration and 
should be expected to control higher derivative corrections 
to the Born-Infeld action. For recent discussion on the geometry 
of corrections to brane actions see \cite{bsw, raams}.  
An effective mass can be 
defined by $ m^2 = E^2 - p^2 $.
For the leading order solution one has $ m = \sqrt{1 + r^4}$.
With this effective
mass, the formulae for the energy and momentum look like
the standard ones of special relativity.

For the leading large $N$ solution
\begin{equation}
  \alpha =  -2 r^3  { ( 1 + r_0^4 )^{(1/2)} \over ( 1 + r^4  )^{(3/2)} }
\end{equation}
This shows that the proper acceleration starts at a small
value  $\alpha  \sim { 1 \over r_0 } $ at the initial position $r_0$, which is
 taken to be much larger
than $1$. It grows to order $1$ at  $ r \sim  r_0^{2/3} $
 and reaches a maximum of $ \alpha \sim r_0^2$ when $ r = 1 $.
It  drops back to order $1$ at $ r \sim r_0^{-2/3}$
and  finally to zero when $ r =0 $.
Near the extrema of  the phase space curves
obtained from $ \cLo$,
where ${ds \over dr}$ is infinite, the acceleration
and proper acceleration continuously approach infinity.
This suggests that the full stringy degrees of freedom
are important, or at least we need some information
about the nature of the terms in the effective action
involving proper accelerations and higher derivatives.

For the cases of spin half, and the Lagrangian $\cLt$,
 we can study the acceleration numerically.
For $ \cLt$, with $ U_0 < \sim C^{1/4}$, the
proper acceleration behaves similarly to
that for the solutions to the zeroth order equations of motion coming from
 $\cLz$, reaching large values
around $ r =1 $. Somewhat surprisingly, for very large $U_0$,  we
find that the acceleration remains small (less than $1 $ in
magnitude) along the trajectory. This is another indication that
for such high $U_0$ the $1/N$ corrected Lagrangians behave very
differently from the leading order ones. The fact that the 
proper acceleration remains small suggests that there may be interesting 
time-dependent physics in string theory  which is non-perturbative in the 
$ 1/N$ expansion but  can be reliably described by the Born-Infeld 
actions,  nelecting higher derivative corrections and 
massive string modes. 

For the $ 1/C$ corrected Lagrangian, $ \cLo $,  we find that there
are regions of phase space where the effective mass becomes
imaginary. Numerical studies show that as we follow the trajectory
of a large $D2$-brane down to small radius in the space
parameterised by $(r,s^2)$, we have a minimum radius and then the
curve continues along increasing $s^2$ and approaches infinite
radius as it asymptotes to a finite upper $s^2 = s_E^2 < 1$. The
acceleration  and proper acceleration approach zero as $s_E$ is
approached. The effective mass remains real at the extremum but
becomes imaginary near $s_E$.

For $ \cLo $ the effective mass squared is
\begin{eqnarray}\label{effmassqL1}
m^2 &\equiv &  E^2 - p^2
    = U^2  \bigl( 1 +  {1 \over 3C}  ( \gamma^2 - U^{-2} )^2  \nnm \\
 && \quad
  - { 1 \over 36 C^2 }  ( \gamma^2 - U^{-2} )^2( 15 \gamma^4 - 16 \gamma^2 +
  2 \gamma^2 U^{-2} - U^{-4}) \bigr)
\end{eqnarray}
The region of interest is at large $U$.
Assuming large $U$ and large $ \gamma $, the coefficient of
$U^2$ in the above expression is
$ 1 + { \gamma^4 \over 3C } - {5 \gamma^8 \over 18 C^2 }$.
This changes sign at $ \gamma^4 = C/2 $.
Note that this value of $\gamma $ is larger than the
large $C$ location of the extremum $ \gamma \sim \sqrt {  2 C \over 5 }  $,
consistent with the numerical evidence that the
extremum is reached before the tachyonic behaviour of the mass squared.
For fixed initial conditions, there is also a tachyon
appearing with $ \cLt$  but this also appears at a larger speed
than the extremum seen with $ \cLo$.

As discussed previously, there are two possible classical
evolutions after the extremum is reached. One involves bouncing
back along the original path, with a discontinuity in velocity.
The other is to bounce back along the branch with increasing
$s^2$. Here we are finding that the first bounce does not
encounter the tachyonic region whereas the second does. The
immediate neighbourhood of the extremum is free from the tachyon
but involves infinite proper accelerations. The correct string 
description may require string field theory.

Plots of the effective mass as a function of $ (r, s ) $ continue
to show the radial mode becoming tachyonic in certain regions, for
$ \cLt$ as well as for the spin half Lagrangian.
Following the classical trajectory of the radius as a function of
$ \gamma $ for $\cLt$, zero radius is reached at finite speed. The
effective mass becomes imaginary at some point along the classical
trajectory if $ U_0 > \sim C^{1/4}$.
The effective squared mass can be plotted as a function of
$(r,s )$ for the spin half case. For $r =0 $ or small, it becomes tachyonic
around $ s = 0.6 $.

A summary of the previous comments is that there is a class 
of perturbative classical paths which  are
modified by small $ 1/C$ corrections. They describe the 
collpase of branes starting at rest at  some  $ U_0 $ which
corresponds to $ \gamma < \sim  C^{1/4}$. In the leading Lagrangian 
$ \cLz $ the limiting $ \gamma \sim C^{1/4}$  are reached for 
$ U_0 \sim  C^{1/4}$. The $ 1/C $ corrections relevant to such paths 
 can be computed
analytically along the lines of Section (4.3) and (4.4). 
 The existence of
interesting features such as the extremum we found with $ \cLo$
requires finite $N$ investigations, which we initiated with the
spin half case.
If we find such extrema in the finite $N$ case, understanding the
complete picture requires information about higher orders in
time derivatives, which should be obtained from string theory.
This is because the proper acceleration diverges at these extrema.
Alternatively we may go beyond the framework of effective actions
and study the problem in string field theory. It may also be
interesting to explore what happens to such extrema when one
considers simple higher derivative Lagrangians which are designed
to put an upper limit on the proper acceleration \cite{nfls}.

\subsection{The physics of the tachyon}

A heuristic argument can be used to predict the existence of the
tachyonic effective mass which we found. Imagine an observer
sitting at the north pole of the collapsing spherical D2-brane, at
$ \Phi_3 = \hat R $. In the rest frame of this observer, the
north-south axis is Lorentz-contracted, so that the 2-brane looks
more like the surface of a pancake. The spherical brane has zero
net D2-charge, and can be thought as having positive charge at the
north pole and negative charge at the south pole. This follows
since the D2-charge density is proportional to $ [ \Phi_1, \Phi_2
] \sim ~ i  \Phi_3$, which is positive at the
north pole and negative at the south pole. Locally then, our
observer at the north pole sees themselves at rest on a D2-brane
with an anti-D2 brane approaching at high speed. We know that the
D2-anti-D2 system has a tachyon when the separation is close to
the string scale. The observer also sees a density of magnetic
flux which is a distribution of D0-charge, but D0-D2 systems also
have a tachyon. It would be interesting to make a more quantitative
connection between this picture and the tachyonic effective mass
obtained from $\cLo, \cLt$ and the spin half case, in order to
understand better this \lq\lq pancake tachyon\rq\rq.
The insights from \cite{baka,bahu,mywi} which discuss
 brane-anti-brane systems in the presence of motion and flux  
 may be useful.    Another
situation where tachyonic behaviour of a radial brane variable has
been found recently is described in \cite{kut}.

It might be argued that the tachyon is an artefact of the
symmetrised trace prescription, which is not the correct
supersymmetric non-abelian DBI theory. It has been shown  that the
symmetrised trace prescription must be corrected, based on BPS
energy formulas \cite{hata, stt, sevwij}. However,
we find it likely that these tachyonic features will survive with
these corrections.
 Indeed it can be argued that $\cLo$ is not modified. 
 The form $ Str  ( \alpha \alpha )^n  =  N ( C^n +
 a_1 ( n-1) C^{n-1}  + n(n-1)(n-2) ( an + b ) C^{n-2} +\dots  )   $
 follows from general arguments
 which will hold true even when the symmetrised trace prescription
 is replaced by something more complicated such as weighting
 different symmetry patterns with different coefficients.
 For example, we know that the first correction of the form $C^{ (n- 1) } $
 must vanish for $n = 0$ and $1$ because for these
 values of $n$ we can simply evaluate the traces and
 check that the leading term $C^n$ is accurate.
 For $n=1$, we have $ tr ( \alpha_i \alpha_i )= N C$.
 Assuming that the $n$-dependent coefficients are sufficiently
nice functions of $n$, it would follow that
 the first correction is unchanged by these corrections. 
 The failure of the symmetrised trace prescription
 starts at $tr(F^6)$ which translates here into $ tr ( \alpha_i\alpha_i )^3 $.
 Once the structure of the $ tr ( F^8 )$ terms are known
 the numbers $ a , b $ appearing in the coeffient
 of $ C^{n-2}$ can be determined, and will likely
 be different from the values $ 7 , -1$ of (\ref{Deqn2}). 
 We have checked that the effective mass
 to order $1/C^2$ can become imaginary at large $ \gamma $
 for {\it any }  choice of $a, b $. It is still possible to
 argue that the tachyon seen in the $1/N$ expansion
 is really an artefact of the failure of the
 $1/N$ expansion itself, and that the correct supersymmetric
 non-abelian Born-Infeld at finite $N$ will not allow such
 tachyonic behaviour. An important future direction is to develop
 a proof (which does not look at all obvious)
along these lines which disposes of the tachyon,
 or to develop a more concrete quantitative formulation of the physical
nature of the tachyon as outlined above. The expectation that the
$1/N$ expansion captures some qualitative features of the finite
$N$ physics suggests that the latter avenue will be more fruitful.
We have also seen that the symmetrised trace prescription does
lead to formulae for the energy which correctly capture the
expected relativistic limit $ s =1 $  in the spin half case.


\section{Summary and outlook}

A large $N$ approximation to the non-abelian zero-brane action of
Myers for time-dependent fuzzy sphere configurations gives
equations for the radius which agree  with the expected dual
picture in terms of a D2-brane with a magnetic flux. These
equations have the same form as those for a relativistic particle
with a position dependent mass.
 The $1/N$ corrections to the
zero-brane action give rise to modifications
of these equations. We studied the energy function
for these modified equations in detail.

The simplest solution  of interest at large $N$, with Lagrangian $
\cLz$,  describes a D2-brane, with initial radius $R \gg l_s $
 collapsing to zero radius.  Classically this
solution can be patched with an expanding solution at zero radius,
but the velocity is discontinuous at that point. The zero brane
description allows us to deduce that the collapsing radius can be
trusted to distances less than the string length. However the
speed at very small  radius for such a large initial membrane is
close to the speed of light. At these speeds, higher orders in the
$1/N$ expansion of the zero brane action become important. Another
feature to bear in mind is that the proper acceleration becomes
large along the classical trajectory. This raises an interesting
question of whether the form of the time-dependent solution can be
argued to be unchanged throughout the trajectory from large $R$
down to zero, perhaps using arguments along the lines of
\cite{larus}.

Using the first order in $1/N$ corrected Lagrangian $ \cLo$, we
showed that
 for a large range of initial conditions
$ R > l_s $ ($r > 1$ in dimensionless variables), the classical
path obtained from $ \cLo $ is qualitatively similar to that
obtained from $\cLz$. However for initial conditions which allow $
\gamma$ to reach near   $ C^{1/4}$ along the path, the path
described in phase space encounters a minimum at non-zero speed.
This conclusion can be reached from the analysis of the energy as
a function of $r$ and $s$, where $r$ is the dimensionless radius
and $s$ the dimensionless velocity. The contours of constant
energy starting from zero velocity  at some large radius cannot be
continued to zero radius, but rather have a minimum. At the
minimum, $ {
\partial E \over \partial s } = 0 $, while  $ s_{min} \ne 0 $.
Such minima do not occur in simple mechanical systems where $ E =
{ m \dot x^2 \over 2 }  + V(x) $. This occurs in our equations
because of the peculiar mixing of the coordinate and velocity in
the energy function. At the extremum, the velocity is close to the
speed of light. As $ N \rightarrow \infty $ this extremum recedes
to the relativistic barrier $s =1 $, which is why we do not see it
in the leading large $N$ approximation. The bounce at the extremum
is somewhat peculiar, it involves a discontinuity in the velocity
and also an ambiguity in the choice of trajectory after the
bounce. These features are made clear by a local analysis of the
energy function in the neighbourhood of the singularity. We
discussed the quantum mechanical behaviour of the system near this
local singularity, and argued that quantum mechanics would provide
probabilities for the two trajectories. The discussion is not
complete since the equations we are trying to quantise have a
non-linear dissipative nature, hinting that the correct quantum
treatment requires extra degrees of freedom, such as those that
naturally occur in the string set-up when we go beyond the low
energy effective action approach and include higher string
excitations.

Another consequence of the fact that the velocity at the extremum
is close to the speed of light is that the $1/N$ correction term
to the Lagrangian is comparable to the leading term for such
velocities. This means that the $1/N$ approximation is breaking
down in the neighborhood of the extremum. This motivated an
analysis of higher orders in the $1/N$ expansion. At the second
order in the $1/N$ expansion we found that the extremum
disappears, while it reappears again at the third order.

This led us to study the exact Lagrangian for finite $N$. In the
case $N=2$, we found that the extremum does not exist, but for
general $N$ the existence or otherwise of such extrema is a very
interesting open question, which could be addressed using the
techniques developed in this paper. If such extrema exist then we
will have a new mechanism in string theory for a brane bounce
which is intimately related to the non-Abelian structure of
D-brane actions.

The technical computations of the $1/N$ corrections require care
in dealing with the symmetrised trace. This symmetrised trace
leads to the evaluation of certain $SO(3)$ invariants. We
presented two approaches to the calculation. One proceeded by
evaluating on highest weights. The other used a diagrammatic
approach for re-ordering the various terms in the symmetrised
product. This latter approach makes interesting connections with
knot theory, which may provide a source of new techniques and
ideas in this context.

Further computations with $\cLo$  included a perturbative
computation of the time taken to collapse from an initial radius
to some final radius. This perturbative approach is best suited to
initial and final radii which are not too far apart. We also
computed the energy as a function of momentum giving an
interesting deformation of the dispersion relation for a massive
relativistic particle, derived directly from the non-Abelian
structure of the zero brane action in string theory. Whereas many
string-inspired deformations of relativistic dispersion relations
have been discussed, this deformation is the first that follows
directly from the non-abelian nature of stringy D-brane actions.

In Section 7, we gave a detailed discussion of regimes of validity
for the different approximations, taking care to use the proper
acceleration rather than the ordinary acceleration as a measure of
whether higher derivative corrections are important. Quite
generically, we found that the generalised effective mass defined
by $ \sqrt{E^2 - p^2}$ became tachyonic in certain regions of
phase space close to the speed of light. We discussed the physics
of this ``pancake tachyon'' as associated with a geometry 
which looks very similar to system of brane and anti-brane. We 
gave arguments against attributing the tachyon to the 
inadequacy of the 
symmetrised trace prescription used in the non-abelian 
Born-Infeld.

Here we have considered the simple situation of a spherical
D2-brane collapsing from a dual D0-point of view using the fuzzy
2-sphere construction. It will be interesting to explore similar
brane collapse phenomena for higher branes using higher
dimensional fuzzy spheres
\cite{gkp,clt,sphdiv,horam,kimur,bag,dcp}.
This may be a new  way to explore cosmological bounces 
in a braneworld context developing works such as \cite{bqrtz}, with the 
additional ingredient of the non-abelian Born-Infeld 
 action.   Position dependent
effective masses 
have  been considered in \cite{questka, dha} and
earlier references therein. The leading order
Lagrangian $ \cLz $ is a simple example of a relativistic system
with position dependent mass. An effective mass can be defined
using $m^2 = E^2 - p^2 $ for the higher order Lagrangians. We
found tachyonic behaviour in certain regions of phase space close
to the speed of light.   This problem of
the collapsing brane may also have similarities with gravitational
collapse of thin shells \cite{crisol}. Another direction is 
to look for a gravitational background which corresponds to the 
time-dependent
system of large $ N$ zero branes, and find a spacetime interpretation for 
some of the features we have found such as the bounces 
and the tachyonic effective masses.


\bigskip

\noindent{\bf Acknowledgements}: We thank
David Berman, Clifford Johnson, Costis Papageorgakis,
Radu Tatar, Gabriele Travaglini and
 Angel Uranga for discussions, and Jos\'e
Figueroa-O'Farrill for the use of his files for drawing chord
diagrams. Thanks to Simon McNamara and Costis Papageorgakis 
for pointing out a missing combinatoric factor in equation (118).

\end{document}